\def\kms{\nobreak\mbox{$\;$km\,s$^{-1}$}}
\def\fm{\hbox{$.\!\!^{\rm m}$}}
\def\la{\mathrel{\hbox{\rlap{\hbox{\lower4pt\hbox{$\sim$}}}\hbox{$<$}}}}
\def\ga{\mathrel{\hbox{\rlap{\hbox{\lower4pt\hbox{$\sim$}}}\hbox{$>$}}}}
\font\teneufm=eufm10
\font\seveneufm=eufm7
\font\fiveeufm=eufm5
\newfam\eufmfam
\textfont\eufmfam=\teneufm
\scriptfont\eufmfam=\seveneufm
\scriptscriptfont\eufmfam=\fiveeufm
\def\frak#1{{\fam\eufmfam\relax#1}}

\documentstyle[psfig]{cupconf}

\title[$H_0$ from Type Ia Supernovae]{{\boldmath $H_0$}
       from Type Ia Supernovae}
\author[G.A. Tammann, A. Sandage \& A. Saha]{
 G.\ls A. T\ls A\ls M\ls M\ls A\ls N\ls N$^1$,\ns
 A. S\ls A\ls N\ls D\ls A\ls G\ls E$^2$ \and\ns A. S\ls A\ls H\ls A$^3$}
\affiliation{$^1$Astronomisches Institut der Universit\"at Basel \\
Venusstrasse 7, CH-4102 Binningen, Switzerland\\[\affilskip]
$^2$The Observatories of the Carnegie Institution of Washington \\
813 Santa Barbara Street, Pasadena, CA 91101\\[\affilskip]
$^3$National Optical Astronomy Observatories \\
950 North Cherry Ave., Tucson, AZ 85726}

\begin{document}
\maketitle

\vspace*{-7cm}
\rightline{A Decade of HST Science} 
\rightline{eds. M. Livio, K. Noll, \& M. Stiavelli}
\rightline{(Cambridge: CUP)}
\vspace*{6.0cm}

\begin{abstract}
The Hubble diagrams in $B$, $V$, and $I$ of a complete sample of 35
SNe\,Ia with $(B-V)<0.06$ and $1200 < v \la 30\,000\kms$ have a
scatter of only $0\fm1$, after small corrections are applied for
differences in decline rate $\Delta m_{15}$ and color $(B-V)$. The
tightness of the Hubble diagrams proves blue SNe\,Ia to be the best
``standard candles'' known. Their absolute magnitudes $M_{B,V,I}$ are
calibrated by eight SNe\,Ia with Cepheid distances from
$HST$. Combining this calibration with the appropriate Hubble diagrams
yields a large-scale value of $H_0=58.5\pm6.3$ at the 90-percent
confidence level.

   The Hubble diagram of SNe\,Ia has so small scatter that it seems
feasable to determine $\Lambda$ ``locally'', i.e. within $z\la0.12$,
once 100-200 SNe\,Ia with good photometry will be available. Such a
local determination would minimize evolutionary effects and K-term
corrections. 

   Clusters of galaxies have provided useful Hubble diagrams through
brightest cluster members, TF distances, and D$_n-\sigma$ and
fundamental plane distances, but with significantly more scatter
($\sigma=0\fm2-0\fm3$) than SNe\,Ia. The zeropoint calibration of
these Hubble diagrams is an additional problem, which is aggravated by
the high weight of any adopted distance of the Virgo cluster and by
selection effects of clusters with only few well studied members. 
If a Virgo cluster modulus of $(m-M)=31.60\pm0.20$ is adopted for
calibration, --- a value which is well secured by Cepheids, SNe\,Ia, and
the TF method, and which also agrees with less definitive distances
from the globular cluster luminosity function, novae, and the
D$_n-\sigma$ method, --- one finds $H_0=55\pm5$. The reasons are
explained why some authors have found higher values from clusters. 

   The determination of $H_0$ from field galaxies is beset by
selection effects of magnitude-limited samples (Malmquist
bias). Authors who have properly allowed for bias have consistently
obtained $H_0\approx 55\pm5$ within $v\la5000\kms$ based on the TF and
other methods. The calibration rests only on Cepheids, independent of
any adopted Virgo cluster modulus.

   Giving highest weight to SNe\,Ia it is concluded that $H_0=58\pm6$. 
\end{abstract}

\section{Introduction}
%
$HST$ has brought an enormous progress in determining extragalactic
distances by providing 26 Cepheid distances to late-type galaxies. This
is of paramount importance not only for the determination of the
Hubble constant $H_0$, but increasingly also for the physical
understanding of individual galaxies whose linear sizes, luminosities,
masses, radiation densities etc. depend on distance. Unfortunately the
progress is confined to late-type galaxies, the distances of
early-type galaxies having profited only indirectly.

   An overview of the Cepheid distances from $HST$ is given in
Table~1. Cepheids carry by far the largest weight for the
foundation of the extragalactic distance scale. Their zeropoint is
based on an adopted LMC modulus of $18.50$, which is confirmed by
various distance indicators to within $\pm0\fm10$. Cepheid distances are
uncontroversial to a large extent with possibly remaining small
metallicity effects discussed in Section~2.5.

\begin{table}
\begin{center}
\caption{Cepheid distances from $HST$.}
\label{tab:1}
\begin{tabular}{llc}
\noalign{\bigskip}
 No. of Galaxies & Authors & No. of SNe\,Ia \\
\hline
18 Cepheid distances & Freedman 2000          & 1 \\
1 Cepheid distance   & Tanvir et~al. 1995     & 1 \\
7 Cepheid distances  & Saha et~al. 1999, 2000a & 8 \\
\end{tabular}
\end{center}
\end{table}

   The 26 galaxies with Cepheid distances from $HST$, all beyond the
Local Group, yield a mean value of $H_0 = 65\pm4$
[km\,s$^{-1}$\,Mpc$^{-1}$]. Yet these galaxies lie within $v=1200\kms$
which is too local for this determination having any cosmic
significance.

   An additional distance indicator is therefore needed which can be
calibrated by means of the available Cepheids and which carries the
distance scale out to $\ga10\,000\kms$, i.e. well beyond the
influence of peculiar and streaming motions. This distance indicator
must be {\em proven\/} to be reliable and its intrinsic dispersion
must be known in order to control selection effects
(cf. Section~4).

   Many distance indicators have been proposed, but the proof of their
reliability is extremely difficult. The demonstration that they can
reproduce a limited number of Cepheid distances, which carry
themselves individual errors of up to $\sim0\fm2$, is not good enough,
and the internal dispersion remains ill defined. 
Just because various distance indicators are said to agree amongst
themselves is no proof that they form a correct distance scale.
If they have similar intrinsic dispersion and if each is not corrected
for systematic effects of observational bias errors, their agreement
is spurious.

   The only satisfactory way to prove potential distance indicators to
be useful for the determination of $H_0$ is the demonstration that
they define a linear relation of slope $0.2$ (corresponding to linear
expansion) in the Hubble diagram out to $\ga 10\,000\kms$. The scatter
about this line provides in addition the intrinsic dispersion of
the method if proper allowance is made for observational errors and
for the influence of peculiar motions which, however, at
$10\,000\kms$, are negligible for all practical purposes.

   This requirement of reliable long-range distance indicators is
very well met by supernovae of type Ia (SNe\,Ia), and they can be
calibrated by Cepheids. SNe\,Ia offer therefore the optimum route to
$H_0$ and are discussed in Section~2. Cluster distances follow in
Section~3; they also define a useful Hubble diagram, but their
zeropoint calibration is still under debate. The most difficult and
least satisfactory way to $H_0$ by means of distances of field
galaxies is discussed in Section~4. The conclusions are given in
Section~5.

\section{{\boldmath $H_0$} from SNe\,Ia}
%
This Section is the result of an $HST$ project for the luminosity
calibration of SNe\,Ia that also includes L.~Labhardt,
F.D.~Macchetto, and N.~Panagia.
The collaboration of J.~Christensen, B.R.~Parodi, H.~Schwengeler, and
F.~Thim at different stages of the project is acknowledged. We thank
also the many collaborators who work behind the scene at the Space
Telescope Science Institute for their continued support.

\subsection{The sample}
From a parent population of 67 SNe\,Ia with known $B$ and $V$ at
maximum, $(B_{\max}-V_{\max})<0.20$ 
(In the following $(B-V)$ for short),
and $v^{\rm CMB}\la 30\,000\kms$ (to minimize
cosmological effects; galaxies with $V_{220} > 3000\kms$ being
corrected for the local $630\kms$ motion with respect to the CMB) a
sample of 44 SNe\,Ia was selected which fulfill the additional
conditions: (1) having occurred after 1985 to ensure photometric
quality, and (2) $v_{220}>1200\kms$ (after correction for Virgocentric
infall).

   The color cut in $(B-V)$ is justified in Fig.~1 where
{\em all\/} SNe\,Ia after 1985 are plotted. Their color distribution
is sharply peaked. Some of the 10 red objects with $(B-V)>0.20$
probably have high internal reddening, others are known to have low
expansion velocities (like SN\,1986G) or quite peculiar spectra (like
SN\,1991bg, 1992K). The latter may define special subclasses of
SNe\,Ia and clearly should be excluded from a homogeneous set of
SNe\,Ia. Their exclusion is in any case indicated as long as none of
the calibrating SNe\,Ia is of this type, which is not the case
(Section~2.4).
\def\floatwidth{0.6\textwidth}
\begin{figure} 
  \centerline{\psfig{file=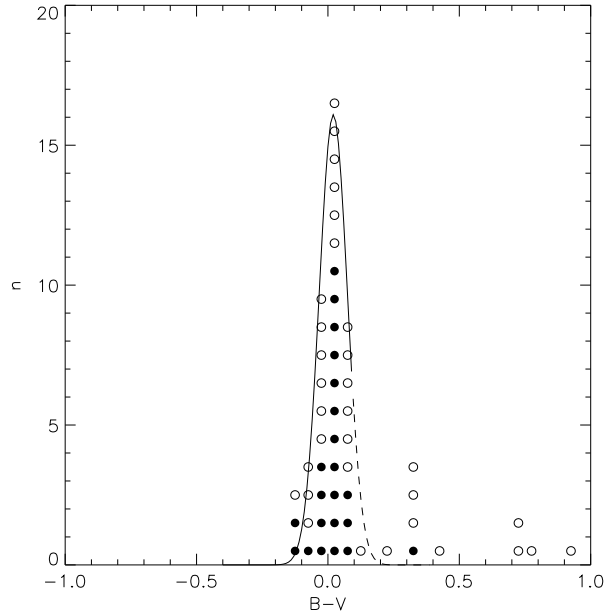,width=\floatwidth}}
\caption{The color distribution of all known SNe\,Ia after 1985 with
  $v<30\,000\kms$. Open symbols are for more distant SNe\,Ia. The
  binned intervals embrace $\Delta (B-V)=0\fm05$. A Gaussian fit to
  all SNe\,Ia with $(B-V)\le0\fm10$ gives $<\!B-V\!> = 0.020,
  \sigma_{\rm B-V}=0\fm053$.}
\label{fig:1}
\end{figure}

   Seven SNe\,Ia of the blue sample have $0.06 < (B-V) < 0.20$. All
seven are underluminous and five of them lie in the inner parts of
spiral galaxies. They are therefore suspected to suffer mild
absorption in their parent galaxies. Their exclusion is permissible
because none of the calibrating SNe\,Ia is as red.

   Two SNe\,Ia of the blue sample, i.e. SN\,1991T
(Phillips et~al.\ 1992) and SN\,1995ac (Garnavich et~al.\ 1996),
had peculiar spectra during their early phases. SN\,1991T has long
been suspected to be overluminous, but a recent Cepheid distance
(Saha et~al.\ 2000a) suggests this overluminosity to be only
marginal. SN\,1995ac lies, however, significantly above the Hubble
line and is certainly overluminous. Both objects are excluded here.

   The remaining 35 blue SNe\,Ia constitute our ``fiducial sample''. They
are listed in Parodi et~al.\ (2000). 
Their spectra, as far as available, are Branch-normal.
For 29 SNe\,Ia also $m_{I}(\max)$ is known.

\subsection{The color of blue SNe\,Ia}
The mean color of 16 SNe\,Ia of the fiducial sample, that have
occurred in early-type galaxies, is $<\!\!B-V\!\!> = -0.013\pm0.015$
and is identical with the mean color of 9 {\em outlying\/} SNe\,Ia in
spirals. We therefore take this as the {\em true\/} mean intrinsic
color of SNe\,Ia. The mean intrinsic color derived by
Phillips et~al.\ (1999) from different assumptions is bluer by
$\sim0.03$. This is irrelevant for the following discussion as long as
the calibrating SNe\,Ia and the remaining objects of the fiducial
sample conform with the adopted zeropoint color. (If our adopted mean
color was too red, any additional absorption would equally affect the
calibrators and the fiducial sample and have no effect on the derived
distances). Indeed, after the eight calibrating SNe\,Ia (Section~2.4)
are individually corrected for absorption they exhibit a mean
color of $<\!\!B-V\!\!> = -0.009\pm0.015$, i.e. only insignificantly
different from our adopted mean intrinsic color. Also the 10 remaining
SNe\,Ia of the fiducial sample, which lie in the inner parts of
spirals or whose position within their parent spirals is poorly known,
are redder by a negligible amount of only $0\fm012\pm0\fm016$.

   The scatter in $(V-I)$ is somewhat larger ($\sigma_{V-I}=0\fm08$)
than in $(B-V)$, but here again the mean color of $<\!\!V-I\!\!> =
-0.276\pm0.016$ is the same for the SNe\,Ia in E/S0 galaxies and
in spirals as well as for the calibrating SNe\,Ia.

\subsection{Second-parameter correction}
The 35 SNe\,Ia of the fiducial sample define very tight Hubble
diagrams in $B$, $V$, and $I$ for the range $1200 < v \la
30\,000\kms$ (cf. Parodi et~al.\ 2000; their Fig.~3), the scatter being
only $\sigma_{B}=0\fm21$, $\sigma_{V}=0\fm18$, and
$\sigma_{I}=0\fm16$. This proves their usefulness as ``standard
candles''.

   Although the Hubble diagrams of SNe\,Ia in $B$, $V$, and $I$ are
tighter than for any other known objects, they still contain systematic
effects. As suspected early on by some authors and pointed out again
by Phillips (1993) the peak luminosity of SNe\,Ia correlates with
the decline rate. Phillips introduced $\Delta m_{15}$, i.e. the
decline in magnitudes during the first 15 days after $B$ maximum, as a
measure of the decline rate. Indeed the residuals of the SNe\,Ia of the
fiducial sample correlate with $\Delta m{15}$ (Fig.~2a). The
relation is roughly $\delta M \propto 0.5 \Delta m_{15}$ in all three
colors, slow decliners being brighter.

\def\floatwidth{0.45\textwidth}
\begin{figure} \medskip
  \centerline{\psfig{file=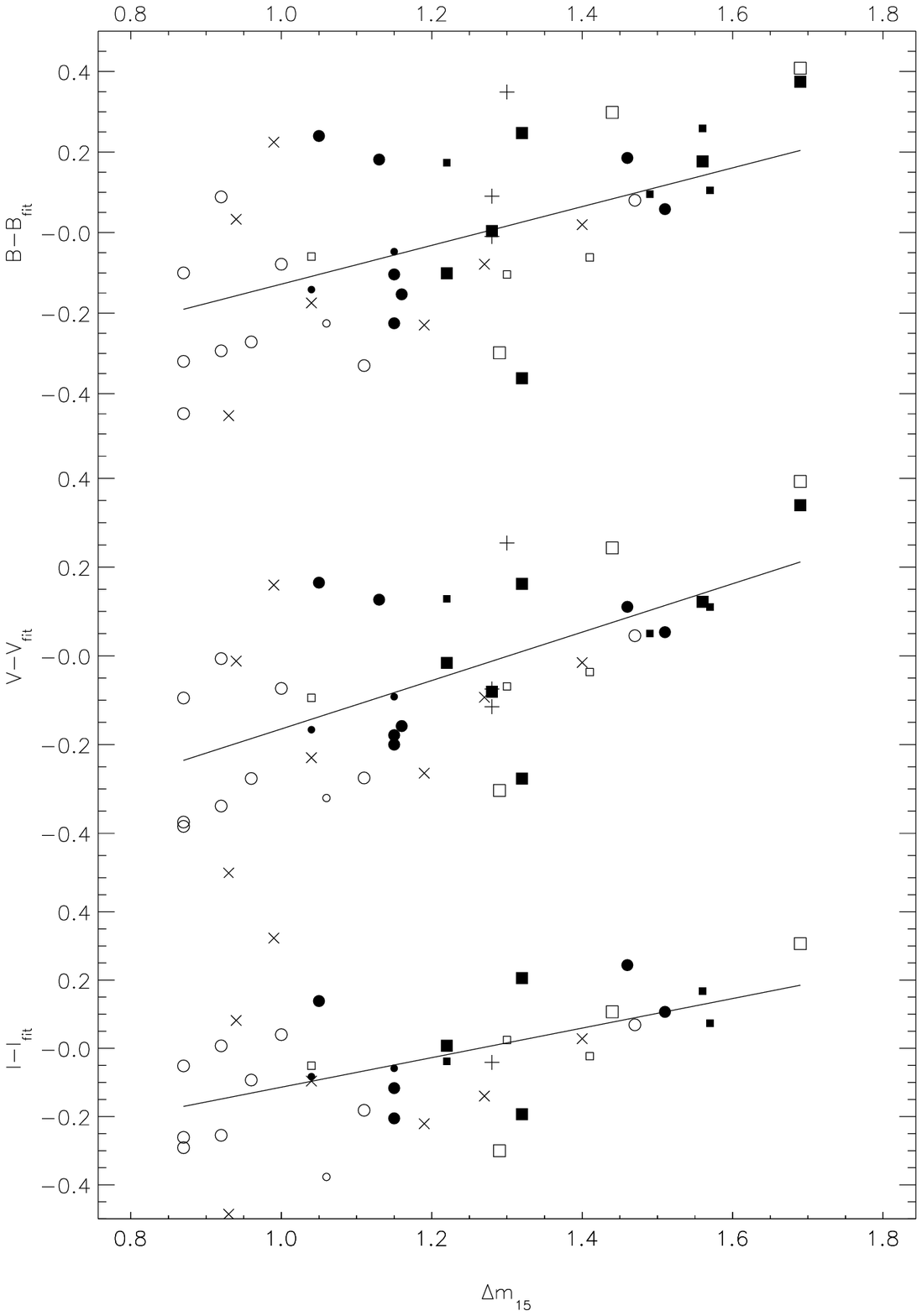,width=\floatwidth}
                \hspace*{0.05\textwidth}
\def\floatwidth{0.463\textwidth}
              \psfig{file=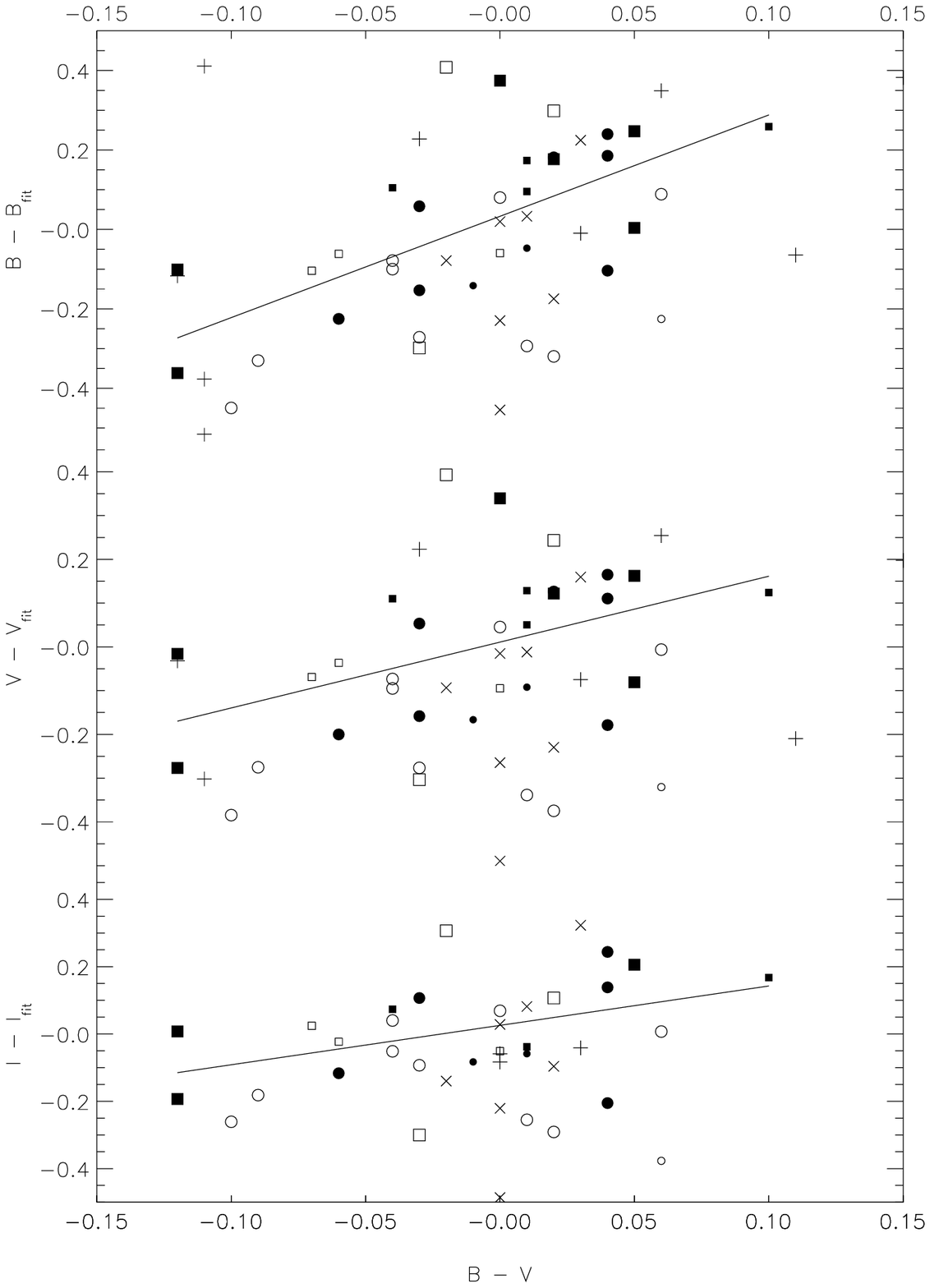,width=\floatwidth}}
\caption{{\em Left panel:} Relative magnitudes (i.e. residuals from
              the Hubble line) for the SNe\,Ia of the fiducial sample
              in function of the decline rate $\Delta m_{15}$. Circles
              are SNe\,Ia in spirals, squares in E/S0 galaxies. Open
              symbols are SNe\,Ia with $1200 < v < 10\,000\kms$,
              closed symbols are for more distant SNe\,Ia. Small
              symbols are SNe\,Ia whose observations begin eight days
              after $B$ maximum or later. Neither the SNe\,Ia before
              1985 with known $\Delta m_{15}$ (shown as crosses) nor
              the seven blue, but reddened SNe\,Ia (shown as X's) are
              considered for the weighted least-squares solutions
              (solid lines).
         {\em Right panel:} Relative absolute magnitudes
              (i.e. residuals from the Hubble line) for the SNe\,Ia of
              the fiducial sample in function of their color
              $(B-V)$. Symbols as in the left panel.}
\label{fig:2}
\end{figure}

   In addition the luminosities of SNe\,Ia correlate with the color
$(B-V)$ (Tammann 1982; Tripp 1998). The proportionality factor
between the magnitude residual $\delta M$ and $\Delta (B-V)$ decreases
from $\sim\!\!2.6$ in $B$ to $\sim\!\!1.2$ in $I$ for the fiducial
sample, blue SNe\,Ia being brighter (Fig.~2b). This
dependence is significantly shallower than for absorption by standard
dust. There are indeed strong reasons (Saha et~al.\ 1999) to believe
that the dependence of luminosity on color is an intrinsic effect of
SNe\,Ia.

   The ``second parameters'' $\Delta m_{15}$ and $(B-V)$ are
orthogonal to each other. A simultaneous fit of the residuals $\delta
M$ in function of $\Delta m_{15}$ {\em and\/} $(B-V)$ is therefore
indicated. In that case the fiducial sample yields (Parodi et~al.\ 2000)
\begin{eqnarray}\label{eq:1}
\!\!\!\!\!\!\!\!\!\!\!\!\!\!\!\!\delta M_B^{\rm corr}&=& 0.44_{\pm0.13}\,(\Delta
m_{15}\!-\!1.2)\!+\!2.46_{\pm0.46}\,[(B\!-\!V)\!+\!0.01]\!-\!28.40_{\pm0.16},\,\sigma_B=0.129\\
\label{eq:2}
\!\!\!\!\!\!\!\!\!\!\!\!\!\!\!\!\delta M_V^{\rm corr}&=& 0.47_{\pm0.11}\,(\Delta
m_{15}\!-\!1.2)\!+\!1.39_{\pm0.40}\,[(B\!-\!V)\!+\!0.01]\!-\!28.39_{\pm0.14},\,\sigma_V=0.129\\
\label{eq:3}
\!\!\!\!\!\!\!\!\!\!\!\!\!\!\!\!\delta M_I^{\rm corr}&=& 0.40_{\pm0.13}\,(\Delta
m_{15}\!-\!1.2)\!+\!1.21_{\pm0.43}\,[(B\!-\!V)\!+\!0.01]\!-\!28.11_{\pm0.17},\,\sigma_I\,=0.122
\end{eqnarray}

   It is interesting to note that the residuals $\delta M$ correlate
also with the Hubble type (SNe\,Ia in early-type galaxies being
fainter) and marginally so with the radial distance from the galaxy
center, but that these dependencies disappear once the apparent
magnitudes are corrected for $\Delta m_{15}$ and $(B-V)$.

   The apparent magnitudes $m_{B,V,I}^{\rm corr}$ of the 35 SNe\,Ia,
corrected for $\Delta m_{15}$ and color by means of
equations~(\ref{eq:1})\,-\,(\ref{eq:3}), define Hubble diagrams as shown in
Fig.~3. {\em Their tightness is astounding.} The Cerro Tololo
collaboration, to whom one ows 70 percent of the photometry of the
fiducial sample, quote a mean observational error of their
$m_{\max}$-values of $\sim\!0\fm10$ and of their colors $(B-V)$ of
$\sim\!0\fm05$. This alone would suffice to explain the observed
scatter of $\sigma_m=0\fm12-0\fm13$. An additional error source are
the corrections for Galactic absorption which were adopted from
Schlegel et~al. (1998). In fact, if one excludes the nine SNe\,Ia
with large Galactic absorption corrections ($A_V>0\fm2$) the scatter
decreases to $0\fm11$ in all three colors. Two important conclusions
follow from this. (1) If the total observed scatter of the Hubble
diagrams is read vertically as an effect of peculiar motions, a
generous upper limit is set of $\Delta v/v=0.05$, which holds for the
range of $3500\la v \la 30\,000\kms$. The (all-sky)
distance-dependent variation of $H_0$ must be even smaller. (2) If
on the other hand the scatter is read horizontally and if allowance is
made for the observational errors of the apparent magnitudes (and for
any peculiar motions) one must conclude that the luminosity scatter of
blue SNe\,Ia, once they are homogenized in $\Delta m_{15}$ and color,
is smaller than can be measured at present. With other words, they are
extremely powerful standard candles.

   It is obvious that if one can determine the absolute magnitude of a
few (nearby) SNe\,Ia this offers what we believe to be a definitive
route to determine the large-scale value of $H_0$.

\subsection{The luminosity calibration of SNe\,Ia}
As stated before, $HST$ has provided Cepheid distances to nine
galaxies which have produced 10 SNe\,Ia. Excluding the
spectroscopically peculiar SN\,1991T in NGC\,4527 (Saha et~al.\ 2000a)
leaves eight galaxies with nine SNe\,Ia. They are listed in
Table~2.

\begin{table}
\caption{Absolute $B$, $V$, and $I$ magnitudes of blue SNe\,Ia
  calibrated through Cepheid distances of their parent galaxies.}
\label{tab:2}
\begin{center}
\begin{minipage}{\textwidth}
\scriptsize
\begin{tabular}{llcccccccl}
\noalign{\medskip}
 SN & \multicolumn{1}{c}{Galaxy} & $(m$-$M)^0$ & ref. &
 $M_{B}^{0}$ & $M_{V}^{0}$ & $M_{I}^{0}$ & $\Delta m_{15}$ \\
 (1) & (2) & (3) & (4) & (5) & (6) & (7) & (8) \\
\hline
1895\,B  & NGC\,5253  & 28.01\,(08) & 2 &
   $-19.54\,(22)$ &  \dots         & \dots          & \dots      \\
1937\,C  & IC\,4182   & 28.36\,(09) & 1 &
   $-19.56\,(15)$ & $-19.54\,(17)$ & \dots          & 0.87\,(10) \\
1960\,F  & NGC\,4496A & 31.04\,(10) & 3 &
   $-19.56\,(18)$ & $-19.62\,(22)$ & \dots          & 1.06\,(12) \\
1972\,E  & NGC\,5253  & 28.61\,(08) & 2 &
   $-19.64\,(16)$ & $-19.61\,(17)$ & $-19.27\,(20)$ & 0.87\,(10) \\
1974\,G  & NGC\,4414  & 31.46\,(17) & 4 &
   $-19.67\,(34)$ & $-19.69\,(27)$ & \dots          & 1.11\,(06) \\
1981\,B  & NGC\,4536  & 31.10\,(05) & 5 &
   $-19.50\,(14)$ & $-19.50\,(10)$ & \dots          & 1.10\,(07) \\
1989\,B  & NGC\,3627  & 30.22\,(12) & 6 &
   $-19.47\,(18)$ & $-19.42\,(16)$ & $-19.21\,(14)$ & 1.31\,(07) \\
1990\,N  & NGC\,4639  & 32.03\,(22) & 7 &
   $-19.39\,(26)$ & $-19.41\,(24)$ & $-19.14\,(23)$ & 1.05\,(05) \\
1998\,bu & NGC\,3368 & 30.37\,(16) & 8 &
   $-19.76\,(31)$ & $-19.69\,(26)$ & $-19.43\,(21)$ & 1.08\,(05) \\
\hline
\multicolumn{4}{r}{mean (straight, excl. SN\,1895\,B)} &
   $-19.57\,(04)$ & $-19.56\,(04)$ & $-19.26\,(06)$ & 1.06\,(05) \\
\multicolumn{4}{r}{mean (weighted,excl. SN\,1895\,B)}  &
   $-19.55\,(07)$ & $-19.53\,(06)$ & $-19.25\,(09)$ & 1.08\,(02) \\
\multicolumn{4}{r}{$M^{\rm corr}(\Delta m_{15}=1.2; (B-V)=-0.01)$} &
   $-19.48\,(07)$ & $-19.47\,(06)$ & $-19.19\,(09)$ & 1.08\,(02) \\

\end{tabular}
References --- (1) Saha et~al.\ 1994 (2) Saha et~al.\ 1995 (3) Saha
et~al.\ 1996b (4) Turner et~al.\ 1998 (5) Saha et~al.\ 1996a (6) Saha
et~al.\ 1999 (7) Saha et~al.\ 1997 (8) Tanvir et~al.\ 1995.
\end{minipage}
\end{center}
\end{table}

   The weights of the individual values of $M$ in Table~2
are quite different due to the different quality of the SNe\,Ia light
curves and of the Cepheid distances, but they are also relatively
strongly affected by the corrections for internal absorption. The
estimated individual errors are compounded and carried on to determine
the total weights. This procedure is much safer than to exclude single
SNe\,Ia for the one or the other reason.

   The adopted mean absolute magnitudes in Table~2 agree
fortuitously well with explosion models, ejecting about
$0.6\,\frak{M}_{\odot}$ of $^{56}$Ni, by H{\"o}flich \& Khokhlov (1996) for
equally blue SNe\,Ia (cf. also Branch 1998).

   The $HST$ photometry for the Cepheids in the galaxies listed in
Table~2 (except NGC\,4414) has been re-analyzed by
Gibson et~al.\ (2000). For 114 Cepheids in common with
Saha et~al.\ (1994, 1995, 1996a,b, 1997, 1999)
they find a brighter photometric zeropoint by
$0\fm04\pm0\fm02$, which is as satisfactory as can be expected from
the subtle photometry with WFPC-2. On the other hand a recent check on
the zeropoint by A.~Saha suggests that it should become
fainter by $0\fm02$.

   Gibson et~al.\ (2000) have added Cepheids which were reduced
only with the photometric ALLFRAME package. Many of the additional
Cepheids had also been detected by us, but were discarded because of
what we considered to be insuperable problems such as poor light
curves or excess crowding.  
With their additional Cepheids Gibson et~al.\ (2000) have derived a
distance modulus of NGC\,5253 that is $0\fm39$ smaller than listed in
Table~2. Their reduction is unlikely for us because it would
imply a very faint tip of the red-giant branch. For the remaining
galaxies in Table~2 they suggest a mean decrease of the
distance moduli by $0\fm11\pm0\fm03$. This would lead to an increase
of $H_0$ by 5\,-\,6 percent. We do not consider this possibility
pending independent confirmation of the ALLFRAME Cepheids.

\subsection{The value of $H_0$}
Combining the weighted mean absolute magnitude $M_{BVI}^0$ of SNe\,Ia
from Table~2 with the observed Hubble diagrams in $B$, $V$,
and $I$ of the fiducial sample, corrected only for Galactic
absorption, leads immediately to a mean value of
$H_0(BVI)=58.3\pm2.0$ (internal error) (cf. Parodi et~al.\ 2000). The
three colors $B$, $V$, and $I$ give closely the same results.

   However, the calibrating SNe\,Ia lie necessarily in late-type
galaxies (because the parent galaxies must contain Cepheids), and
these SNe\,Ia are therefore expected to be somewhat more luminous than
their counterparts of the fiducial sample which lie in galaxies of all
Hubble types (cf. Section~2.3).
Because of the correlation between Hubble type and decline rate
$\Delta m_{15}$, the calibrators should have also slower decline rates
than average. This is indeed the case. Consequently
the calibrators and the fiducial sample should be homogenized as to
$\Delta m_{15}$, i.e. the corrected magnitudes $M_{BVI}^{\rm corr}$
from Table~2 should be compared with the corrected Hubble
diagram in Fig.~3.
It may be noted that the correction for variations in color $(B-V)$ has
here no net effect because the calibrators and the fiducial sample
have identical mean colors.

\def\floatwidth{0.5\textwidth}
\begin{figure} \medskip
  \centerline{\psfig{file=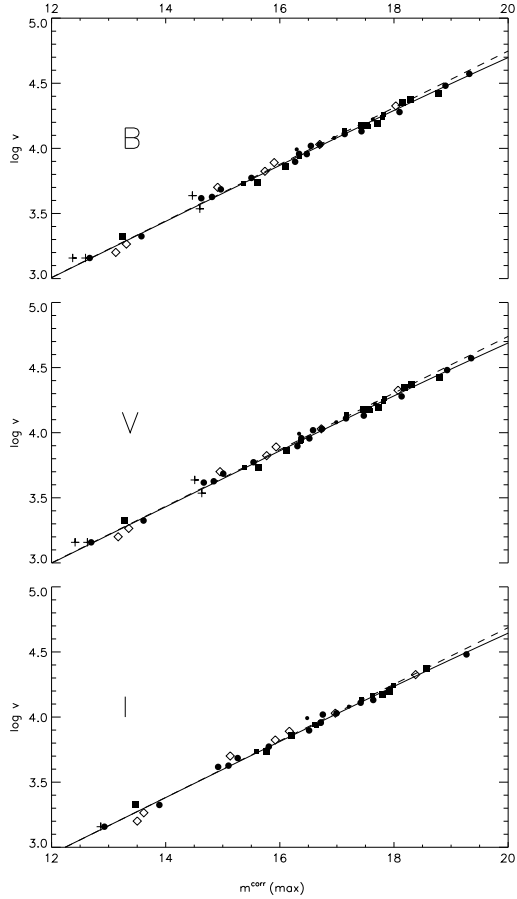,width=\floatwidth}}
\caption{The Hubble diagrams in $B$, $V$, (and $I$) for the 35 (29)
  SNe\,Ia of the fiducial sample with magnitudes $m^{\rm corr}$,
  i.e. corrected for decline rate $\Delta m_{15}$ and color $(B-V)$
  (equations~\ref{eq:1}\,-\,\ref{eq:3}). Circles are SNe\,Ia in
  spirals, squares in E/S0 galaxies.
  Small symbols are SNe\,Ia whose observations begin eight
  days after $B$ maximum or later. Solid lines are fits to the data
  assuming a flat universe with $\Omega_{\rm M}=0.3$ and
  $\Omega_{\Lambda}=0.7$; dashed lines are linear fits with a forced
  slope of 0.2 (corresponding approximately to $\Omega_{\rm M}=1.0$ and
  $\Omega_{\Lambda}=0.0$). Not considered for the fits are the SNe\,Ia
  before 1985 and the seven SNe\,Ia with $0.06 < (B-V) < 0.20$ that
  are suspected to be reddened. They are shown as diamonds after
  absorption correction; their inclusion would have nil effect on the
  fit.}
\label{fig:3}
\end{figure}

   An exact comparison should allow for the fact that cosmological
effects on the Hubble diagram are non-negligible at
$\sim\!\!30\,000\kms$.
Three different model universes are therefore fitted to the data as
illustrated in a differential Hubble diagram (Fig.~4):

\def\floatwidth{0.65\textwidth}
\begin{figure} \medskip
  \centerline{\psfig{file=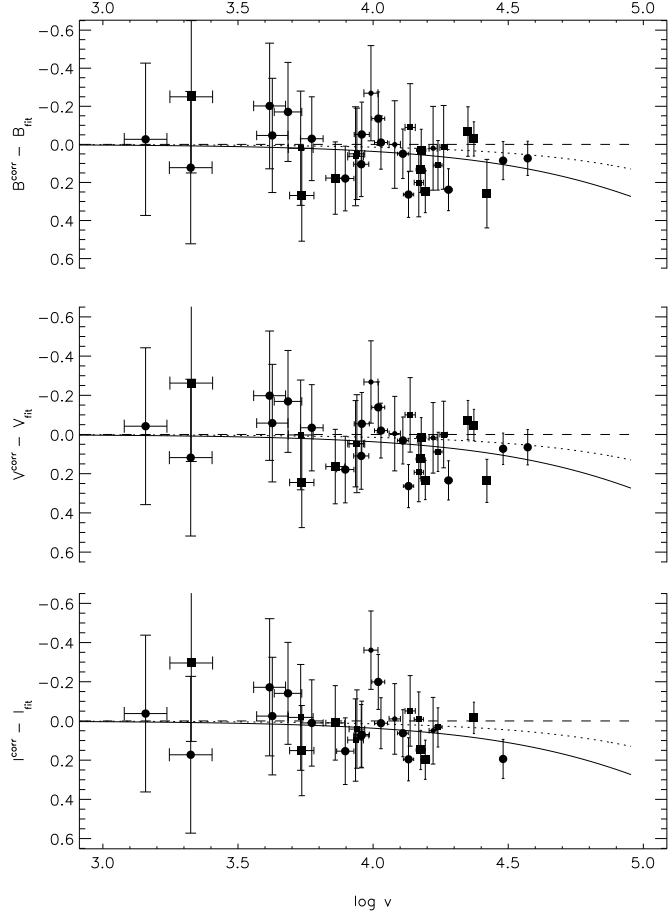,width=\floatwidth}}
\caption{Differential Hubble diagrams ($m^{\rm corr}-m_{\rm fit}$) vs
  $\log v$ in $B$, $V$, (and $I$) for the 35 (29) SNe\,Ia of the
  fiducial sample. Symbols as in Fig.~3. The dashed line is for a flat
  cosmological model with $\Omega_{\rm M}=1.0$ and
  $\Omega_{\Lambda}=0.0$; the theoretical apparent magnitudes $m_{\rm
  fit}$ correspond to this model. The full line is for a flat model
  with $\Omega_{\rm M}=0.3$ and $\Omega_{\Lambda}=0.7$; the dotted
  line is for an open universe with $\Omega_{\rm M}=0.2$ and
  $\Omega_{\Lambda}=0.0$.}
\label{fig:4}
\end{figure}

\noindent
1.~ A flat Universe with $\Omega_{M}=1.0$ ($q_0=0.5$; Sandage 1961,
1962). When the magnitudes $m_{BVI}^{\rm corr}$ of the fiducial sample
are fitted to the corresponding Hubble line one obtains, after
inserting  $<\!\!M_{BVI}^{\rm corr}\!\!>$ of the calibrators,
$H_0(B)=60.2\pm2.1$. The values in $V$ and $I$ are very similar (60.1
and 60.0, respectively).

\noindent
2.~ An open Universe with $\Omega_{M}=0.2$ ($q_0=0.1$; Sandage 1961,
1962). The fit is in this case somewhat better, giving
$H_0(BVI)=60.2\pm2.0$.

\noindent
3.~ A flat Universe with $\Omega_{M}=0.3$, $\Omega_{\Lambda}=0.7$
($q_0= -0.55$). This model is favored by {\em
  high-redshift\/} SNe\,Ia (Perlmutter 1998, Riess et~al.\ 1998,
Schmidt et~al.\ 1998, Perlmutter et~al.\ 1999).
The Hubble line is given in this case by  (cf. Carroll, Press, \&
Turner 1992)
\begin{equation}\label{eq:4}
m_{B,V,I}=5\log \left( \frac{c}{H_0} (1+z_1)\int_{0}^{z_1} [(1+z)^2(1+\Omega_Mz)-z(2+z)\Omega_\Lambda]^{-1/2}dz \right) + M_{B,V,I}+25\,.
\end{equation}
This model gives the best fit to the fiducial sample and yields, after
insertion of $<\!\!M_{BVI}^{\rm corr}\!\!>$ from Table~2,
$H_0(B)=61.0\pm2.1$, $H_0(V)=60.9\pm1.8$, and $H_0(I)=60.7\pm2.6$.
A mean value of  $H_0=60.9\pm1.8$ is adopted for the following
discussion.

   The three model Universes can be distinguished with the present
fiducial sample only at the 1$\sigma$-level. But the possibility to
determine $\Lambda$ from rather local SNe\,Ia is interesting in
principle. The advantage would be good multi-color photometry (e.g. in
$B,V,I$), allowing homogenization of the SNe\,Ia in color and good
control of internal absorption, quite small K-corrections, and short
look-back times minimizing evolutionary effects. From 100\,-\,200
SNe\,Ia with good observations and $z < 0.12$ one should obtain a
significant value of $\Lambda$.

\subsection{Discussion}
It should be noted that the second-parameter corrections increase
$H_0$ by only 4.3 percent, i.e. from $58.3$ to $60.9$. This is
supported by the fact that if one restricts the discussion to
SNe\,Ia in spirals, minimizing in this way the difference between the
parent galaxies of the calibrators and of the SNe\,Ia of the fiducial
sample, one obtains $H_0(BVI)=59.1$ independent of any
second-parameter correction. Alternatively, if one considers the 21
SNe\,Ia of the fiducial sample with $\Delta m_{15} \ge 1.3$, whose
mean decline rate of $<\!\!\Delta m_{15}\!\!>=1.08\pm0.02$ is the same
as that of the calibrators, one obtains $H_0(BVI)=59.2$ (Parodi
et~al.\ 2000).

Other authors have derived from part of the fiducial sample and from
all or some of the calibrators in Table~2 values from
$H_0=50\pm3$ (Lanoix 1998) to $H_0=72\pm4$
(Richtler \& Drenkhahn 1999). Intermediate values of
$H_0=63-64\pm2.2$ were found by Suntzeff et~al.\ (1999) and
Phillips et~al.\ (1999), however they base their result on a
significantly steeper $\Delta m_{15}$-luminosity correction than found
here. This would lead to an overcorrection of the larger fiducial
sample used here. If their correction was applied to the present data,
one would obtain $H_0=59.7$ for the SNe\,Ia with $\Delta m_{15} < 1.2$
(n=17), and $H_0=64.6$ for those with $\Delta m_{15} \ge 1.2$ (n=18).
Since seven of the eight calibrators fall into the first category, the
lower value must be more nearly correct. 

   By relying exclusively on the Cepheid distances by Gibson
et~al.\ (2000; cf. Section~2.4), Freedman (2000) was able to push the
solution of Suntzeff et~al.\ (1999) and Phillips et~al.\ (1999) up to
$H_0=68$.  

   Riess et~al.\ (1998) employed a so-called ``Multi Light Curve
Shape'' (MLCS) method to correct simultaneously for Galactic
absorption and the relative SN\,Ia luminosity. The resulting distance
moduli (their Table~10) imply, however, that $H_0$ depends on their
correction parameter $\Delta$, i.e. $<\!\!H_0\!\!>=66.6\pm1.1$ for
$\Delta < -0.20$ (n=9) and $<\!\!H_0\!\!>=61.8\pm1.3$ for $\Delta >
-0.20$ (n=18). Jha et~al. (1999) employed also the MLCS method to
derive $H_0=64\pm7$ without listing $\Delta$-values of individual
SNe\,Ia.  Tripp \& Branch (1999), correcting for $\Delta m_{15}$ and
color $(B-V)$, have obtained $H_0=62 \,(\pm4)$.

   The present result of $H_0=60.2\pm2.1$ is still affected by
external errors. Yet the largest systematic error source of all
distance indicators that depend on an adopted mean luminosity (or
size), i.e. selection effects against underluminous (or undersized)
``twins'', is negligible in the case of blue SNe\,Ia because of their
exceptionally small luminosity dispersion.

   External errors of either sign are introduced (cf. Parodi
et~al.\ 2000) by the photometric zeropoint of the WFPC-2 photometry
($0\fm04$), by the adopted slope of the $\Delta m_{15}$-luminosity
relation ($0\fm02$), by the velocity correction for Virgocentric
infall and by the correction for motion relative to the CMB
($0\fm02$).

   Other errors are asymmetric with a tendency to underestimate
distances. The adopted LMC modulus of $18.50$ for the zeropoint of the
Cepheid PL relation is probably too small by $\sim\!0\fm06$
(eg. Federspiel, Tammann, \& Sandage 1998; Madore \& Freedman 1998;
Feast 1999; Walker 1999; Gilmozzi \& Panagia 1999; Gratton 2000;
Sakai, Zwitsky, \& Kennicutt 2000).
Smaller LMC moduli suggested on the basis of statistical parallaxes of
RR Lyrae stars and red giant clump stars depend entirely on the sample
selection and on the absence of metallicity and evolutionary effects,
respectively. The higher LMC modulus will increase all moduli by
$0\fm06\pm0\fm10$. Incomplete Cepheid sampling near the photometric
threshold always tends to yield too short distances (Sandage 1988a;
Lanoix, Paturel, \& Garnier 1999; Mazumdar \& Narasimha 2000).
The effect is estimated here to be $0\fm05\pm0\fm05$. Stanek \&
Udalski (1999)  have proposed that photometric blends of Cepheids in
very crowded fields lead to a serious underestimate of the
distances. Careful analyses by Saha et~al.\ (2000b) and Ferrarese
et~al.\ (2000) show the effect to be more modest for the Cepheid
distances in Table~2, say $0\fm03\pm0\fm03$ on
average. Absorption corrections of the calibrating SNe\,Ia and those
of the fiducial sample, having identical colors after correction,
enter only differentially. However, if one excludes the nine SNe\,Ia
with large Galactic absorption (Section 2.3) the Hubble line of
Fig.~3 shifts faintwards by $0\fm05$ in $B$. Finally seven
SNe\,Ia were excluded on the suspicion of having some internal
absorption (Section 2.1). If they had been included after being
corrected for absorption, their effect on the Hubble line would be
negligible. If, however, their colors are intrinsic they would shift
the Hubble line by $0\fm02$ towards fainter magnitudes. Combining the
absorption errors it is estimated that the distances of the fiducial
sample are too small by $0\fm05\pm0\fm05$ on average. Finally there is
much discussion of the metallicity effect on Cepheid
distances. Theoretical investigations show this effect to be nearly
negligible (Sandage, Bell, \& Tripicco 1999; Alibert et~al.\ 1999)
Other authors do not even agree on the sign of the
correction. Based on Kennicutt et~al.\ (1998) Gibson et~al.\ (2000) have
concluded that the Cepheid distances in Table~4 are too
small by $0\fm07$ due to variations of the metallicity. A distance
increase of $0\fm04\pm0\fm10$ is adopted as a compromise.

   Adding the various error sources in quadrature leads to a
correction factor of $0.96\pm0.08$ which is to be applied to
$H_0=60.2\pm2.1$. At a 90-percent confidence level one obtains then
\begin{equation}\label{eq:5}
H_0=58.5\pm6.3 .
\end{equation}
If only the Cepheid distances of Gibson et~al.\ (2000;
cf. Section~2.4) had been used, excluding their unlikely value for
NGC\,5253, one would have obtained $H_0=61.6\pm6.6$.

   The standards for the determination of $H_0$ are now set by
SNe\,Ia. In the next Section it will be asked to what extend these
standards can be met by other distance indicators, i.e. how reliable
are they as relative distance indicators and how accurate is their
zeropoint calibration? 

\section{{\boldmath $H_0$} from Cluster Distances}
%
\subsection{The Hubble diagram of clusters}
\subsubsection{Brightest cluster members}
A deep (to $z=$ 0.45) and tight ($\sigma_{m}= 0\fm32$) Hubble Diagram is
that of 1st-ranked E and S0 cluster galaxies corrected for Galactic
absorption, aperture effect, K-dimming, Bautz-Morgan effect, and
cluster richness (Sandage \& Hardy 1973). They define a Hubble line in
the range of $1200<v<30\,000\kms$ of
\begin{equation}\label{eq:6}
  \log v = 0.2\,m_{V_{c}} + (1.359 \pm 0.018); 
  \quad \sigma_{m_{V}}=0.32; \quad n=76
\end{equation}
which is easily transformed into
\begin{equation}\label{eq:7}
\log H_0 = 0.2\,M_V({\rm 1st}) + (6.359\pm0.018) .
\end{equation}

   Weedman (1976) has established a Hubble diagram using the mean
magnitude of the 10 brightest cluster members. The small scatter of
$\sigma_{m_{10}} = 0\fm15$ is not directly comparable with other values
because the magnitudes are defined within a {\em metric\/} diameter
and depend somewhat on redshift.

   Another Hubble diagram of 1st-ranked galaxies has been presented by
Lauer \& Postman (1992). It comprises the complete sample of 114 Abell
clusters with $v<15\,000\kms$. The scatter about the mean Hubble line
amounts to $\sigma_m=0\fm3$.

\subsubsection{The Tully-Fisher (TF) relation}
Dale et~al. (1999) have derived {\em relative\/} $I$-band TF distances
of 52 clusters from an average of 8\,-\,9 members per cluster. The {\em
  mean\/} cluster distances define a Hubble line with a scatter of
only $\sigma_{(m-M)}=0\fm12$, i.e. similar to SNe\,Ia. The clusters
are corrected for {\em differential\/} selection bias, but the data
are still unsuitable to derive absolute distances because with only a
few members per cluster the Teerikorpi cluster incompleteness bias
must be severe (Section~3.4). The same holds for an older sample of 18
clusters by Giovanelli et~al.\ (1997), for which Sakai et~al.\ (2000) have
derived $I$-band TF distances. Sixteen of their clusters with
$1500<v<9000\kms$ and each with 15 studied galaxies on average define
a Hubble line with a scatter of $\sigma_{(m-M)}=0\fm18$, which implies a
scatter of $\sigma>0\fm5$ of the TF modulus of an individual galaxy.

\subsubsection{The D$_n-\sigma$ relation}
Kelson et~al.\ (2000) have derived D$_n-\sigma$ and fundamental plane
(FP) distances of (only) 11 clusters out to $v\!\sim\!10\,000\kms$. The
two resulting sets define Hubble diagrams with a remarkably small
scatter of $\sigma_{(m-M)}=0\fm19$.

\subsubsection{Combining different Hubble diagrams}
A combination of data of brightest cluster galaxies and D$_n-\sigma$
measurements by Faber et~al.\ (1989) have been used to derive cluster
distances relative to the Virgo cluster (Jerjen \& Tammann 1993). When
these are augmented by the relative TF distances of clusters by
Giovanelli (1997) one obtains a Hubble diagram as shown in
Fig.~5. The data scatter by $\sigma_{(m-M)}=0\fm20$ about a
Hubble line. The latter is found by linear regression and implies
\begin{equation}\label{eq:8}
   \log H_{0} = -0.2\,(m-M)_{\rm Virgo} + (8.070 \pm 0.007).
\end{equation}
(Federspiel, Tammann, \& Sandage 1998). The equation is particularly
useful because the calibration of $H_0$ can simply be accomplished by
inserting the Virgo cluster modulus.

   It should be noted that the observed scatter of the above Hubble
diagrams is almost entirely due to intrinsic scatter of the distance
indicators and to measurement errors because the contribution of
peculiar velocities is $\la0\fm1$ as shown by the Hubble diagram of
SNe\,Ia (Fig.~3).

\def\floatwidth{0.65\textwidth}
\begin{figure}
  \centerline{\psfig{file=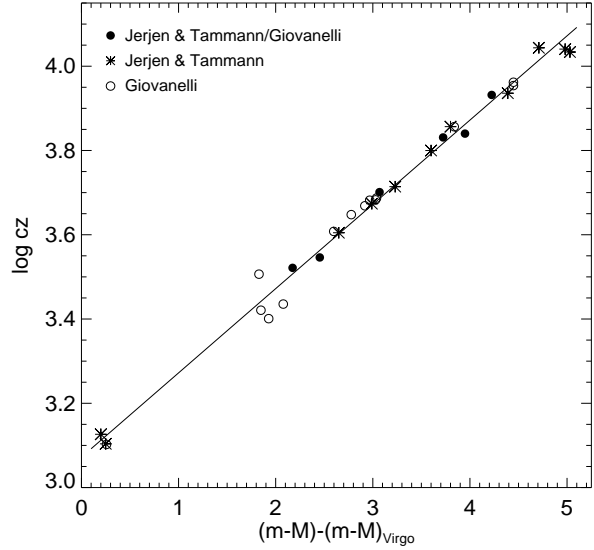,width=\floatwidth}}
\caption{Hubble diagram of 31 clusters with known relative
  distances. Asterisks are data from Jerjen \& Tammann (1993). Open
  circles are from Giovanelli (1997). Filled circles are the
  average of data from both sources.}
\label{fig:5}
\end{figure}

\subsection{Relative distances of local groups and clusters}
The reliability of other distance indicators, which have failed so far
to establish a Hubble diagram, can be tested by comparing their
relative distances of local groups or clusters, specifically between the
Leo\,I group, the Virgo cluster, and the Fornax cluster. It is a most
gentle test because the mean distances from several member galaxies
are compared and the problem of the zeropoint calibration does not enter.

\begin{table}
\caption{Relative distances of Leo\,I, Virgo, and Fornax from various
  distance indicators. (The errors shown are from the quoted sources)}
\label{tab:3}
\begin{center}
\footnotesize
\begin{minipage}{\textwidth}
\begin{tabular}{lcccll}
\noalign{\medskip}
 & \multicolumn{1}{c}{$\Delta(m-M)$} &
   \multicolumn{1}{c}{$\Delta(m-M)$} &
   \multicolumn{1}{c}{$\Delta(m-M)$} & & \\
\multicolumn{1}{c}{Method} &
\multicolumn{1}{c}{Virgo\,--\,Leo\,I} &
\multicolumn{1}{c}{Fornax\,--\,Virgo} &
\multicolumn{1}{c}{Fornax\,--\,Leo\,I} &
\multicolumn{1}{c}{Source} \\
\multicolumn{1}{c}{(1)} & \multicolumn{1}{c}{(2)} &
\multicolumn{1}{c}{(3)} & \multicolumn{1}{c}{(4)} &
\multicolumn{1}{c}{(5)} \\
\hline
Cepheids, SNe\,Ia  & $1.37\pm0.21$ & $0.05\pm0.22$ & $1.42\pm0.10$
                   & Tables~4 and 6 \\
D$_{\rm n}-\sigma$ & $0.92\pm0.32$ & $0.17\pm0.16$ & $1.09\pm0.32$
                   & Faber et~al.\ 1989 \\
FP                 & \multicolumn{1}{c}{$\cdots$} & $\;(0.52\pm0.17)\;$
                   & \multicolumn{1}{c}{$\cdots$} & Kelson et~al.\ 2000 \\
TF ($BVRI$)        & $\;\,(1.95\pm0.23)^1$ & $\!\!\!(-0.40\pm0.10)\;$ &
               \multicolumn{1}{l}{$\,(1.55)^1$} & Schr{\"o}der 1995 \\
vel. ratio$^2$     & $1.29\pm0.31$ & $0.27\pm0.30$ & $1.57\pm0.18$
                   &  Kraan-Korteweg 1986 \\
\hline
 mean: & $1.25\pm0.15$ & $0.15\pm0.12$ & $1.43\pm0.08$ \\
\hline
PNe                & $0.86\pm0.09$ & $0.33\pm0.09$ & $1.19\pm0.10$
                   & Ferrarese et~al.\ 2000 \\
SBF                & $0.88\pm0.07$ & $0.40\pm0.05$ & $1.28\pm0.06$
                   & Ferrarese et~al.\ 2000 \\
                   & $0.93\pm0.05$ & $0.37\pm0.04$ & $1.30\pm0.05$
                   & Tonry et~al.\ 2000 \\
GCLF               & $1.62\pm0.31$ & $\!\!\!-0.35\pm0.14\;$ & $1.27\pm0.31$
                   & Tammann \& Sandage 1999 \\
                   & $1.69\pm0.56$ & $0.14\pm0.07$ & $1.83\pm0.56$
                   & Ferrarese et~al.\ 2000 \\
\end{tabular}
$^1$ From only 3 spirals in the Leo\,I group (Federspiel 1999)

$^2$ Assuming $v_{220}({\rm Leo})=648$, $v_{220}({\rm
Virgo})=1179$, $v_{220}({\rm Fornax})=1338\kms$ (Kraan-Korteweg 1986)
and allowing for a peculiar velocity of Leo\,I of $100\kms$ and of
Fornax of $200\kms$
\end{minipage}
\end{center}
\end{table}

   In Table~3 the modulus differences Virgo\,-\,Leo\,I,
Fornax\,-\,Virgo, and Fornax\,-\,Leo\,I are listed as derived from
various distance indicators. The sources of these differences are
also listed. The first four lines show the differences of the
distance indicators which have passed the Hubble diagram test. The
fifth line gives the relative moduli as derived from the mean
group/cluster velocities. Most of the entries are in statistical
agreement with the weighted means in the sixth line.

There are, however, four entries shown in parentheses, which deviate
significantly from the adopted means. The result of the FP that the Fornax
cluster is more distant by as much as $0\fm5$ than the Virgo cluster
is very unlikely. The TF distance of the Leo\,I group carries little
weight being based on only three galaxies, and the small TF distance
of the Fornax cluster, i.e. $0\fm40$ nearer than the Virgo cluster, is
simply impossible. A compilation (Tammann \& Federspiel 1997) of the
relative distance Fornax\,-\,Virgo, as determined by 30 different
authors, suggests Fornax to be slighty more distant than Virgo in
agreement with the adopted means in Table~3, but all
attempts to determine the TF distance of Fornax have come out with 
suspiciously small values. The discrepancy 
is not sufficiently explained by the fact that even a complete sample
of the Fornax cluster contains only eight full-size spirals suited for
the TF method in addition to 19 small late-type galaxies with large
scatter. The discrepancy remains hence enigmatic. The case shows 
the possibility that even tested distance indicators may fail in
individual cases.

   In the lower part of Table~3 the relative distances are
shown by different authors from planetary nebulae (PNe), the surface
brightness fluctuations (SBF) and of the globular cluster luminosity
function (GCLF). The modulus differences of the PNe are rather
small. In particular the value of Virgo\,-\,Leo\,I is
$0\fm39\pm0\fm17$ smaller than adopted. This unsatisfactory 
result of the PNe does not come as a surprise. The method
rests on the assumption that the luminosity of the brightest planetary
shell, as seen in a bright emission line, is a fixed standard candle
and does not depend on sample size (i.e. galaxy luminosity). This is
not only against the expectation from any realistic luminosity
function, but is also contradicted by observations (Bottinelli
et~al.\ 1991; Tammann 1993; Soffner et~al.\ 1996). Moreover the shell
luminosity is predicted to depend on metallicity and age (M{\'e}ndez
et~al.\ 1993).  

Also the SBFs give a small difference Virgo\,-\,Leo\,I and an almost
certainly too large difference Fornax\,-\,Virgo. If the quoted small
errors are taken at face value the method is incompatible with the
adopted relative position Leo\,I\,-\,Virgo\,-\,Fornax. In any case the
method should be given low weight until its real capabilities are
proved beyond doubt.

   The distance differences of the GCLFs listed in Table~3
have too large errors and differ too much between different
authors, due to different sample selections, that any clear conclusion
could be drawn. In the case of the Virgo cluster it yields a perfect
distance determination (cf. Table~5); in other cases it
gives quite erratic results (Tammann \& Sandage 1999).

\subsection{The distance of the Virgo cluster and of other clusters}
Four galaxies with known Cepheid distances lie in the Virgo cluster
proper, i.e. within the isopleths and the X-ray contours
(cf. Binggeli, Popescu, \& Tammann 1993). As can be seen in
Table~4 they have widely different distances. Three of the
galaxies have been selected from the atlas of Sandage \& Bedke (1988)
on grounds of their exceptionally good resolution; they are therefore
{\em expected\/} to lie on the near side of the cluster. The fourth
galaxy, NGC\,4639, which has a {\em low\/} recession velocity and can
therefore not be assigned to the background, but must be a dynamical
member of the cluster, is more distant by almost $1\fm0$ and must lie
on the far side of the extended cluster. The relative position of the
four galaxies is fully confirmed by their TF distances. The cluster
{\em center\/} must lie somewhere between the available Cepheid
distances, say at $(m-M)=31.5\pm0.3$. Much of the confusion of the
extragalactic distance scale comes from the ill-conceived notion that
the three highly resolved Virgo galaxies could reflect the {\em
  mean\/} distance of the cluster.
\begin{table}
\caption{Distances of the Virgo and Fornax clusters}
\label{tab:4}
\begin{center}
\scriptsize
\begin{minipage}{\textwidth}
\begin{tabular}{llll|llll}
\noalign{\bigskip}
\multicolumn{4}{c}{\normalsize Virgo} & \multicolumn{4}{c}{\normalsize Fornax} \\
 & & & & & & & \\
\multicolumn{1}{c}{Object} & \multicolumn{1}{c}{$(m-M)^0$} &
\multicolumn{1}{c}{Ref.} & \multicolumn{1}{c|}{Remarks} &
\multicolumn{1}{c}{Object} & \multicolumn{1}{c}{$(m-M)^0$} &
\multicolumn{1}{c}{Ref.} & \multicolumn{1}{c}{Remarks} \\
\multicolumn{1}{c}{(1)} & \multicolumn{1}{c}{(2)} &
\multicolumn{1}{c}{(3)} & \multicolumn{1}{c|}{(4)} &
\multicolumn{1}{c}{(5)} & \multicolumn{1}{c}{(6)} &
\multicolumn{1}{c}{(7)} & \multicolumn{1}{c}{(8)} \\
\hline
\multicolumn{8}{c}{Cepheids} \\
\hline
NGC\,4321 & 31.04 & 1 & highly resolved & NGC\,1326A & 31.49 & 7 & \\
NGC\,4535 & 31.10 & 2 & highly resolved & NGC\,1365  & 31.39 & 8 & \\
NGC\,4548 & 31.04 & 3 & highly resolved & NGC\,1425  & 31.81 & 9 & \\
 \multicolumn{1}{c}{\dots} & & & & & & & \\
NGC\,4639 & 32.03 (!) & 4 & normal resolved& & & &\\[1.5ex]
mean: & \multicolumn{3}{l|}{$\sim\!31.5$} &
      & \multicolumn{3}{l}{$31.56\pm0.13$} \\
\hline
\multicolumn{8}{c}{SNe\,Ia} \\
\hline
SN\,1984A & 31.42 & 5 & in NGC\,4419 & SN\,1980N & 31.76 & 5 & in NGC\,1316 \\
SN\,1990N & 32.12 & 5 & in NGC\,4639 & SN\,1981D & 31.51 & 5 & in NGC\,1316 \\
SN\,1994D & 31.27 & 5 & in NGC\,4526 & SN\,1992A & 31.84 & 5 & in NGC\,1380 \\[1.5ex]
mean: & \multicolumn{3}{l|}{$31.60\pm0.30$} &
      & \multicolumn{3}{l}{$31.70\pm0.10$} \\
\hline
\multicolumn{8}{c}{Tully-Fisher Relation} \\
\multicolumn{8}{c}{(complete samples)} \\
\hline
mean (n=49): & $31.65\pm0.25$ & 6 &
             & & $(31.25\pm0.10)$  & 10 &  \\
\hline
overall mean: &
\multicolumn{3}{l|}{$31.60\pm0.20$}&
overall mean: &
\multicolumn{3}{l}{$31.65\pm0.08$}\\
\noalign{\smallskip}
\hline
\end{tabular}
References --- (1) Ferrarese et~al.\ 1996 (2) Macri et~al.\ 1999
  (3) Graham et~al.\ 1999 (4) Saha et~al.\ 1997 (5) see text
  (6) Federspiel et~al.\ 1998; Federspiel 1999 (7) Prosser et~al.\ 1999
  (8) Silbermann et~al.\ 1999 (9) Mould et~al.\ 2000
  (10) Schr{\"o}der 1995; Federspiel 1999

\end{minipage}
\end{center}
\end{table}

   Three well observed SNe\,Ia have appeared in Virgo cluster
members. Their distances in Table~4, corrected for decline
rate $\Delta m_{15}$ and color $(B-V)$, have been calculated from
their individual parameters as compiled by Parodi et~al.\ (2000) and from
the calibration in Table~2.

   The best possible application of the TF method is provided by the
Virgo cluster, because a very deep catalog of the cluster (Binggeli,
Sandage, \& Tammann 1985) allows selection of a {\em complete\/}
sample of all 49 sufficiently inclined cluster spirals. They define a
reliable position and slope of the TF relation in $B$
(Fig.~6b). (It is sometimes argued that $I$-magnitudes
[albeit incomplete!] should be used to minimize the internal-absorption
correction, but the advantage is offset by the steeper slope of the TF
relation at long wavelengths [Schr{\"o}der 1995]).
Combining these data with the excellent calibration of the TF relation
(Fig.~6a), which rests now on 26 galaxies with Cepheid
distances and 2 companions of M\,101, for which a Cepheid distance is
available, yields the TF modulus shown in Table~4.

\def\floatwidth{0.45\textwidth}
\begin{figure} \medskip
  \centerline{\psfig{file=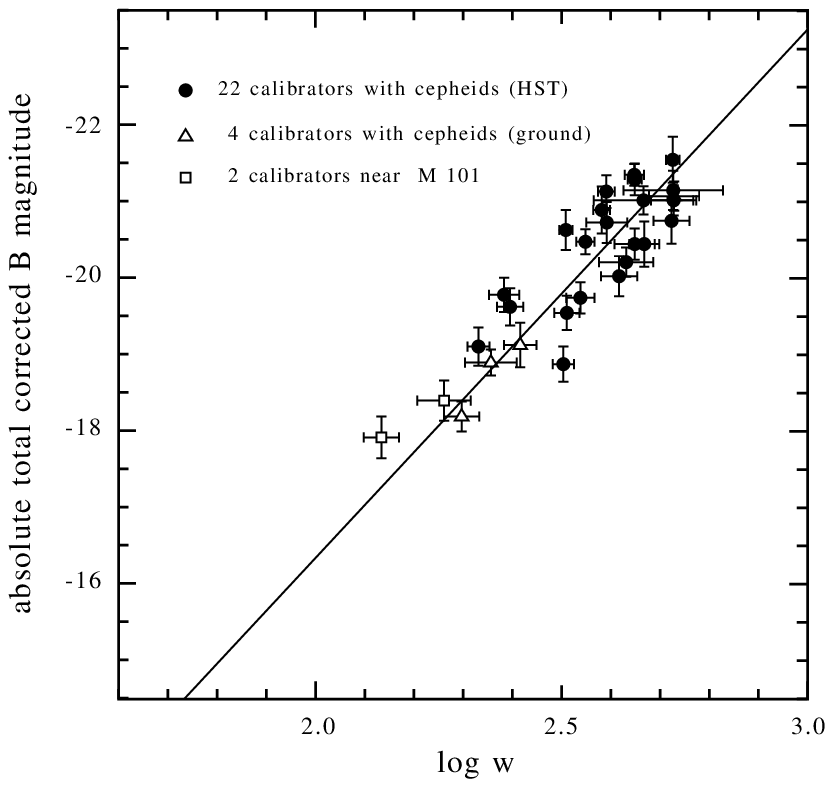,width=\floatwidth}
                \hspace*{0.05\textwidth}
              \psfig{file=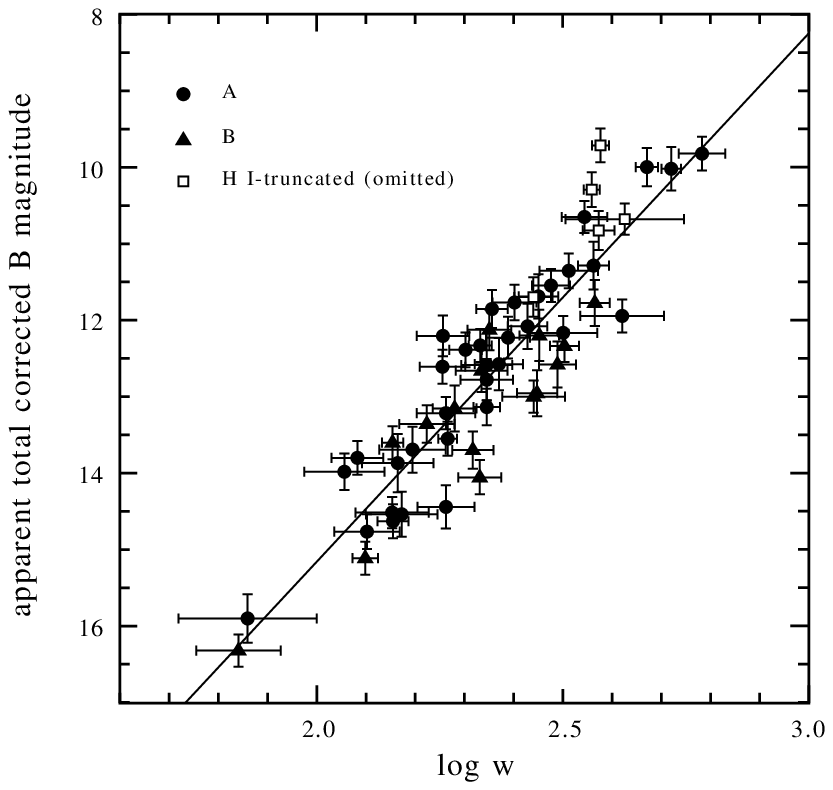,width=\floatwidth}}
\caption{ a) Tully\,-\,Fisher relation of 28 galaxies with
                independently known (Cepheid) distances. -- b)
                Tully\,-\,Fisher relation for a complete sample of 49
                Virgo cluster spirals.}
\label{fig:6}
\end{figure}

   The distance of the Virgo cluster is such an important milestone
for the extragalactic distance scale that the value adopted in
Table~4 should be compared with other distance
indicators. Three distance indicators of early-type galaxies shall be
considered, although their reliability is much less tested.
(1) The peak of the luminosity
function of globular clusters (LFGC) is interesting because its
calibration of the zeropoint rests on the Cepheid distance of M\,31
{\em and\/} in excellent agreement on the RR Lyrae distance of
Galactic globular clusters. (2) Six known novae in Virgo cluster
ellipticals can be compared to the novae in M\,31 whose {\em
  apparent\/} distance modulus can be derived from Cepheids {\em or\/}
from Galactic novae (Capaccioli et~al. 1989). Alternatively the
semi-theoretical zeropoint of novae can be taken from Livio
(1997). (3) The $D_{\rm n}-\sigma$ relation can be applied to E/S0
galaxies and to the bulges of S0\,-\,Sb galaxies. The mean of the two
applications is given here. The zeropoint of the S0\,-\,Sb bulges is
better determined than that of E/S0 galaxies,
because the zeropoint of the latter depends on the assumption that the
only two early-type galaxies of the Leo\,I group lie at the same
distance as its spiral members. The results of the three methods are
compiled in Table~5 and discussed in more detail by Tammann,
Sandage, \& Reindl (2000). The resulting mean distance modulus is in
excellent agreement with the adopted value in Table~4 and is
consistent with the assumption that early-type galaxies and spirals of
the Virgo cluster are at the same distance.

\begin{table}
\footnotesize
\begin{center}
\caption{Additional distance determinations of the Virgo cluster.}
\label{tab:5}
\begin{tabular}{lll}
\noalign{\medskip}
  Method & $(m\!-\!M)_{\rm Virgo}$ & original source \\
\hline
 Globular Clusters    & $31.70 \pm 0.30$ &  Tammann \& Sandage 1999 \\
 Novae                & $31.46 \pm 0.40$ & Pritchet \& van\,den Bergh$\;$1987 \\
 D$_{\rm n} - \sigma$ & $31.70 \pm 0.15$ & Dressler 1987,
                                           Faber et~al.\ 1998 \\
\hline
 mean: & $31.66\pm0.17$ & \\[1ex]
\end{tabular}
\end{center}
\end{table}

   Table~4 shows also the distance modulus of the Fornax
cluster as derived from the three galaxies with Cepheids and three
SNe\,Ia. The individual distances show much less scatter than in the
case of the Virgo cluster; this is obviously a result of the smaller
size and depth of the Fornax cluster. 

   For comparison with other distance indicators it is useful to have
also the Cepheid and SNe\,Ia distance of the Leo\,I group. The
relevant data are set out in Table~6. The adopted mean
modulus is consistent with the GCLF distance ($30.08\pm0.29$; Tammann
\& Sandage 1999) and the surface brightness fluctuation (SBF) distance
($30.30\pm0.06$; Ferrarese et~al.\ 2000).

\begin{table}
\caption{The distance of the Leo\,I group.}
\label{tab:6}
\begin{center}
\footnotesize
\begin{tabular}{llcll}
\noalign{\medskip}
\multicolumn{1}{c}{Method} & \multicolumn{1}{c}{Object} &
\multicolumn{1}{c}{$(m-M)^0$} & \multicolumn{1}{c}{Source} &
\multicolumn{1}{c}{Remarks} \\
\multicolumn{1}{c}{(1)} & \multicolumn{1}{c}{(2)} &
\multicolumn{1}{c}{(3)} & \multicolumn{1}{c}{(4)} &
\multicolumn{1}{c}{(5)} \\
\hline
Cepheids & NGC\,3351 & $30.01\pm0.15$ & Graham et~al.\ 1997 & \\
         & NGC\,3368 & $30.37\pm0.16$ & Tanvir et~al.\ 1995 & \\
         & NGC\,3627 & $30.22\pm0.12$ & Saha et~al.\ 1999 & \\
SNe\,Ia  & 1998bu    & $30.32\pm0.15$ & $m_{BVI}^{\rm corr}(\max)$ +
                                        Table~2 & in NGC\,3368 \\ 
\hline
mean:    &           & $30.23\pm0.07$  &   & \\[1ex]
\end{tabular}
\end{center}
\end{table}

   Finally, it is noted that the distance of the Coma cluster relative
to the Virgo cluster is well determined. The value $\Delta (m-M)_{\rm
  Coma-Virgo}=3.71\pm0.08$ (Tammann \& Sandage 1999) from brightest
cluster galaxies and the TF and D$_n-\sigma$ methods is quite
uncontroversial. With the Virgo modulus in Table~4 one
obtains $(m-M)_{\rm Coma}=35.31\pm0.22$ in excellent agreement with
the independently calibrated GCLF distance from $HST$ (Baum et~al.\ 1997;
cf. Tammann \& Sandage 1999).  

\subsection{The Teerikorpi cluster incompleteness bias}
It is not the place here to discuss the Teerikorpi cluster
incompleteness bias (Teerikorpi 1997 and references therein; Sandage,
Tammann, \& Federspiel 1995) in any detail, because it is well known
that distance indicators with non-negligible internal scatter yield
{\em too small distances\/} if applied to {\em incomplete\/} cluster
samples.

   For illustration the complete sample of 49 Virgo spirals as shown
in Fig.~6b was cut at different apparent-magnitude limits,
and each time the mean TF distance of the remaining sample was
calculated. The mean distance of each sample increases monotonically
with the depth of the cut. The result is shown in
Fig.~7. The true, asymptotic distance is only reached if one
samples $\sim\!4\fm0$ into the cluster.

\def\floatwidth{0.7\textwidth}
\begin{figure} \medskip
  \centerline{\psfig{file=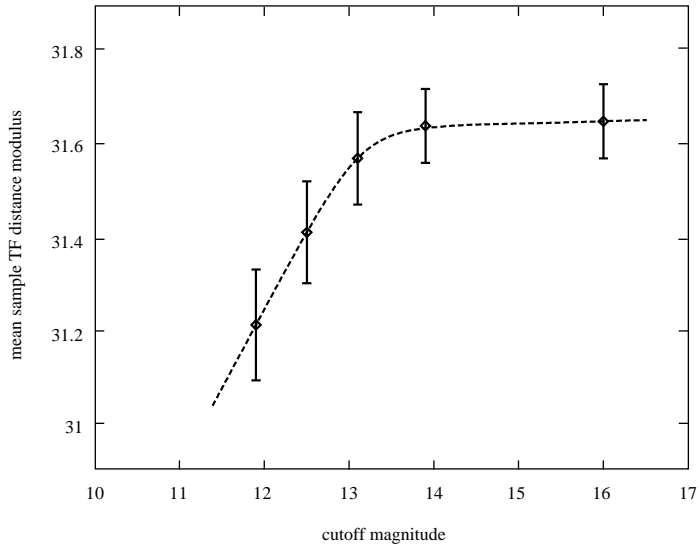,width=\floatwidth}}
\caption{Illustration of the Teerikorpi cluster incompleteness
  bias. If the complete sample of 49 Virgo spirals is cut at levels
  brighter than $m_{B}\approx 13\fm5$ one derives too small TF
  distances. A cut at $m_{B}=12\fm0$ introduces a distance modulus
  error of $0\fm4$ (20 percent in distance).}
\label{fig:7}
\end{figure}

   The bias effect explains why Sakai et~al.\ (2000) have derived too
high a value of $H_0$ by directly applying the TF calibration in
Fig~6a to 15 clusters of Giovanelli et~al.\ (1997). The
shallow cluster samples are far from being complete and are bound to
yield too small cluster distances. In fact the authors have derived 
$<\!\!H_0\!\!>=86$ for the clusters whose distances rest on less than
10 members, while they have found $<\!\!H_0\!\!>=70$ for the clusters
with 10 to 28 members. {\em It is clear that more nearly complete samples
would give still lower values of $H_0$}.

\subsection{The value of $H_0$ from clusters}
The distances of the Virgo, Fornax, and Coma clusters in Section~3.3
yield the absolute magnitudes of three 1st-ranked cluster galaxies
(Table~7).
\begin{table}
\begin{center}
\caption{The absolute magnitude of 1st-ranked cluster galaxies.}
\label{tab:7}
\begin{minipage}{0.75\textwidth}
\begin{tabular}{lrcc}
\noalign{\medskip}
 Cluster & $m_{V_{c}}(1st)^1$ & $(m-M)$ & $M_V(1st)$ \\
\hline
Virgo  &  $8.21$ & $31.60\pm0.20$ & $-23.39\pm0.36$ \\
Fornax &  $8.83$ & $31.65\pm0.08$ & $-22.82\pm0.32$ \\
Coma   & $11.58$ & $35.31\pm0.22$ & $-23.73\pm0.39$ \\
\hline
 mean: & & & $-23.26\pm0.20$ \\
\hline
\end{tabular}
$^1$ From Sandage \& Hardy (1973). A scatter of $0\fm32$ is allowed for.
\end{minipage}
\end{center}
\end{table}
Inserting the mean value into equation~(\ref{eq:7}) gives
\begin{equation}\label{eq:9}
   H_0 = 51 \pm 7 .
\end{equation}

Combining alternatively the cluster distances relative to the Virgo
cluster as given by equation~(\ref{eq:8}) (cf. Fig.~5) with the
Virgo cluster modulus of $31.60\pm0.20$ from Table~4 gives
\begin{equation}\label{eq:10}
   H_0 = 56 \pm 6 .
\end{equation}
The result shows that once the Virgo cluster distance is fixed the
value of $H_0$ has little leeway.

   The relatively large error of the Virgo modulus is due to the
important depth effect of the cluster. The four Cepheid and three
SNe\,Ia distances  do not suffice to sample the cluster in
depth. The situation is aggravated by the fact that the Cepheid
distances of three galaxies are biased because they were selected on
the grounds of high resolution and hence must lie on the near side of
the cluster. If one took unjustifiedly the mean distance of only these
three Cepheids as the true cluster distance one would derive $H_0=72$.

   Fortunately the SNe\,Ia, the most reliable distance
indicators known, confirm the Virgo cluster distance estimated from
Cepheids {\em and\/} the important depth of the cluster. Moreover, the
now very solid Cepheid calibration of the TF relation finds its most
powerful application in the complete sample of Virgo spirals and
corroborates the conclusions from Cepheids and SNe\,Ia.

   Kelson et~al.\ (2000) have derived $H_0=75$ and 80 from the Hubble
diagram based on D$_n-\sigma$ and FP data (Section~3.1.3). They have
chosen certain distances of Leo\,I, Virgo, and Fornax for the calibration.
Had they used the distances given in Table~4 and 6, they would have
found $H_0=60 (\pm6)$ and $65(\pm6)$, instead. This is a clear
demonstration that the D$_n-\sigma$/FP route to $H_0$ is in no way an
independent method, but depends entirely on distances used as
calibrators, and in particular on their distance of the Virgo cluster
which we dispute. The power of $H_0$ from SNe\,Ia, the calibration of
which only depends on Cepheids, bypassing the still controversially
discussed Virgo cluster distance, becomes here evident. 

   Lauer et~al.\ (1998) have tried to calibrate the Hubble diagram of
1st-ranked cluster galaxies of Lauer \& Postman (1992; Section~3.1.1) by
means of the SBF method. For this purpose they have observed the SBF
with $HST$ of four 1st-ranked galaxies at $\sim\!4300\kms$. On the
unproven assumption that the fluctuation magnitude $\overline{m}_I$ of
these very particular objects is the same as in local E/S0 galaxies
{\em and\/} spiral bulges, and adopting a local calibration magnitude
$\overline{M}_I$, which in turn is controversial, they have derived
distances of these four galaxies which they claim are in agreement
with the turnover magnitudes $m_I^{\rm T}$ of the respective
GCLFs. However, their calibrating turnover magnitude $M_I^{\rm T}$
{\em rests entirely on M\,87}, which is known to have a peculiar bimodal
GCLF, {\em and on an adopted Virgo cluster modulus which is $0\fm6$
smaller than shown in Table~4 and 5}. Thus, judging
only from GCLFs their proposed value of $H_0=82$ should be reduced by a
factor of 1.32 to give $H_0=62$. In any case their procedure appears
like a complicated way -- particularly in comparison to SNe\,Ia
-- to transport the Virgo cluster distance into the expansion field at
$v \sim 10\,000\kms$.

\section{{\boldmath $H_0$} from Field Galaxies}
The most difficult and least satisfactory determination of $H_0$ comes
from field galaxies. The difficulty comes from selection effects
(Malmquist bias); the restricted impact comes from the fact that the
method can hardly be carried beyond $\sim\!\!5000\kms$ and hence does
not necessarily reflect the large-scale value of $H_0$.

   The problem of selection effects is illustrated in
Fig.~8b. 200 galaxies of constant space density and with
$v<5000\kms$ were randomly distributed in space by a Monte Carlo
calculation. The ``true'' distances were expressed in velocities. It
has further been assumed that the distance moduli of each galaxy had
also been determined by the TF relation with an intrinsic scatter of
$\sigma_{(m-M)}=\pm0\fm4$. The corresponding Hubble diagram in
Fig.~8b gives the false impression as if the scatter would
increase at larger distances; actually the increasing {\em number\/}
of distant galaxies produces deviations by $\pm2$ or even $\pm3$
sigma. This makes the Hubble diagram ``ugly'' as compared to the one
of SNe\,Ia in Fig.~8a (repeated here from Fig.~3
for comparison), but still the mean Hubble line through the points has
the correct position {\em provided the sample is complete out to a
  given distance limit}. If, however, the sample is cut by an {\em
  apparent-magnitude limit\/} $m_{\rm lim}$ the Hubble line shifts
upwards, resulting in too high a value of $H_0$. With the present,
realistically chosen parameters ($<\!\!M\!\!>=-20.0, \sigma_M=0\fm4,
H_0=60, m_{\rm lim}=13.0$) the overestimate of $H_0$ amounts to
14\%. Samples which are not even complete to a given
apparent-magnitude limit can give much larger systematic errors.

\begin{figure} \medskip
\begin{minipage}[c]{0.48\textwidth}
\psfig{file=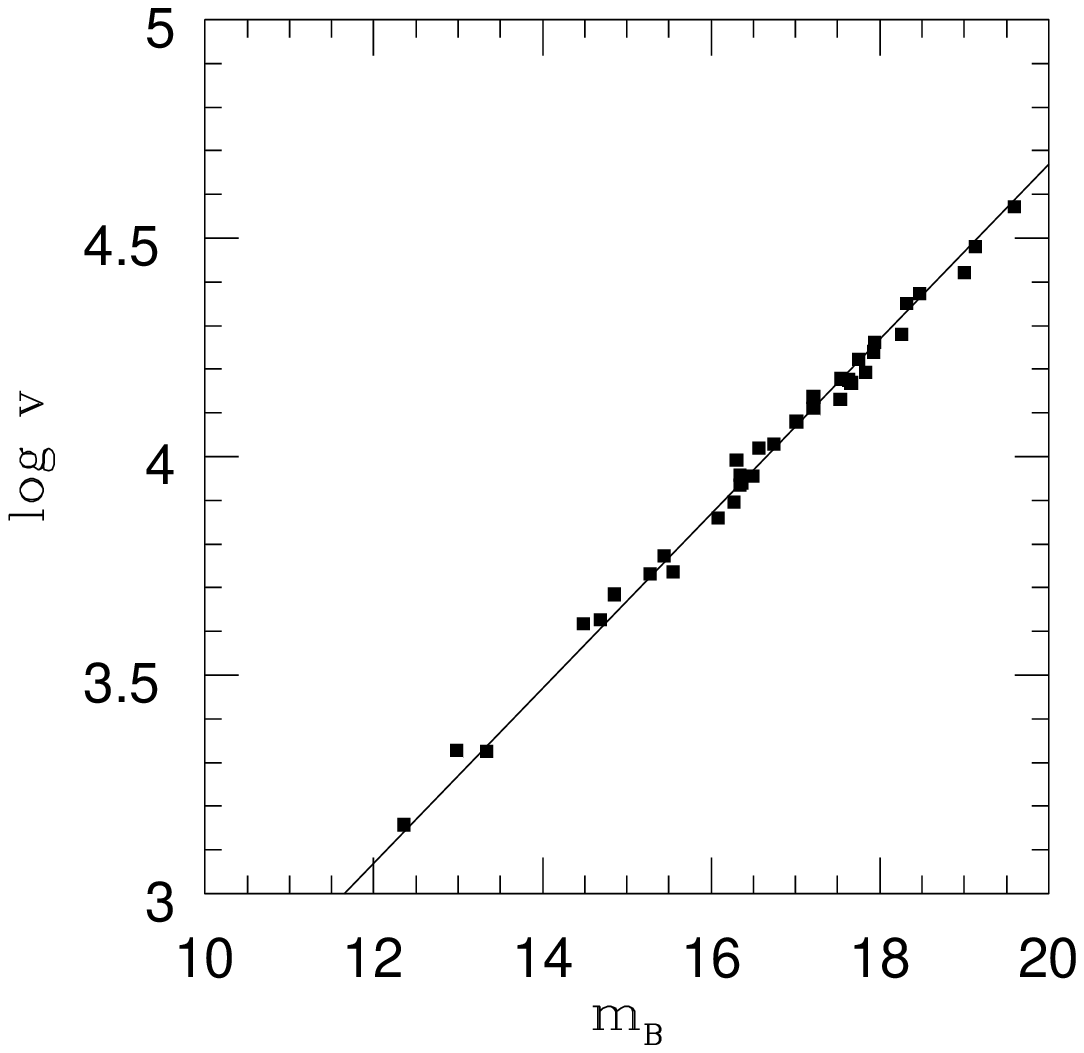,width=\textwidth}
\end{minipage}\nolinebreak 
\begin{minipage}[c]{0.48\textwidth}
\psfig{file=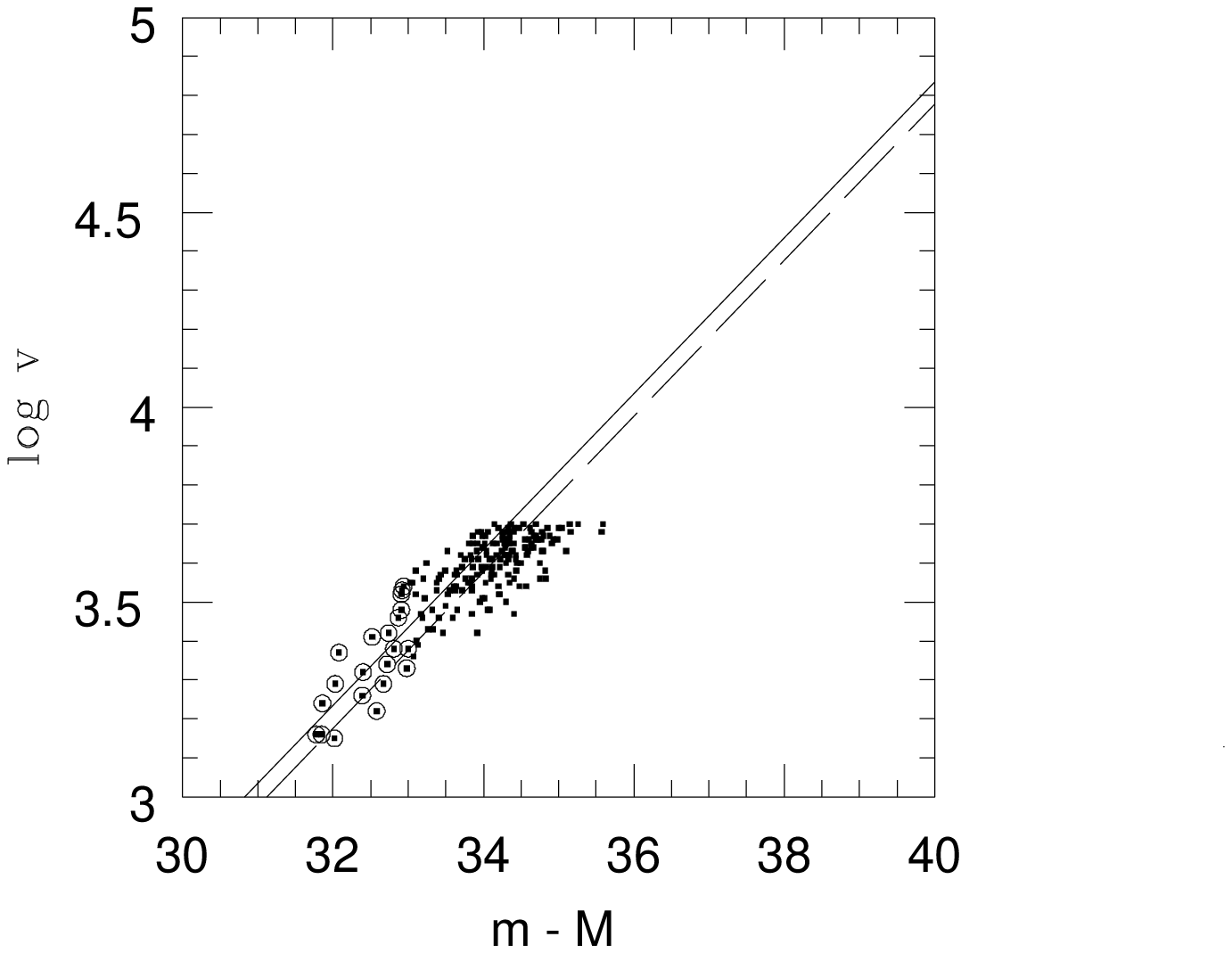,width=1.29\textwidth}
\end{minipage}
\caption{a) The Hubble diagram in $B$ of SNe\,Ia (from Fig.~3; shown
            here only for comparison).
         b) The Hubble diagram of 200 galaxies with
         $v<5000\kms$. Their velocities were assigned by a Monte Carlo
         calculation assuming constant space density. For each galaxy
         it was assumed that its TF distance has been determined
         within an intrinsic scatter of $0\fm4$. For the calibration
         of the abscissa it was assumed that all galaxies have a true
         absolute magnitude of $M=-20.0$ and that $H_0=60$. The dashed
         line is the best fit of the Hubble line for all points. The
         galaxies with {\em apparent\/} magnitude $m<13.0$ are shown
         as large symbols. They define a Hubble line (full line) which
         is $0\fm29$ brighter, corresponding to a too high value of
         $H_0$.}
\label{fig:8}
\end{figure}

   Another illustration of the Malmquist bias is given in the
so-called Spaenhauer diagram (Fig.~9), where 500 galaxies
were randomly distributed in space out to $42\;$Mpc using a Monte
Carlo routine. The galaxies are assumed to scatter by $\sigma_M=2\fm0$
about a fixed mean luminosity of $M=-18.0$. If this distance-limited
sample is cut by a limit in apparent magnitude a sample originates
with very complex statistical properties. In particular the mean
luminosity within any distance interval increases with distance. In
the lower panel of Fig.~9 the increase amounts to $\Delta M=
2\fm6$ which would overestimate $H_0$ by a factor of 3.3 if one were
to force the local calibration of $M=-18.0$ on the most distant
galaxies of the biased sample. For the sake of the argument the
intrinsic dispersion of $\sigma_M=2\fm0$ was chosen here to be
unrealistically large, but the effect is omnipresent in all
apparent-magnitude-limited samples, {\em it always overestimates\/}
$H_0$. Only if $\sigma_M \la 0\fm2$, as in the case of SNe\,Ia, the
Malmquist bias becomes negligible. An additional crux of the bias is
that the observable scatter within any distance interval {\em is always
smaller than the true intrinsic scatter. This has mislead several
authors to assume that $\sigma_M$ is small and hence to underestimate
the importance of bias}.

\def\floatwidth{0.65\textwidth}
\begin{figure}
  \centerline{\psfig{file=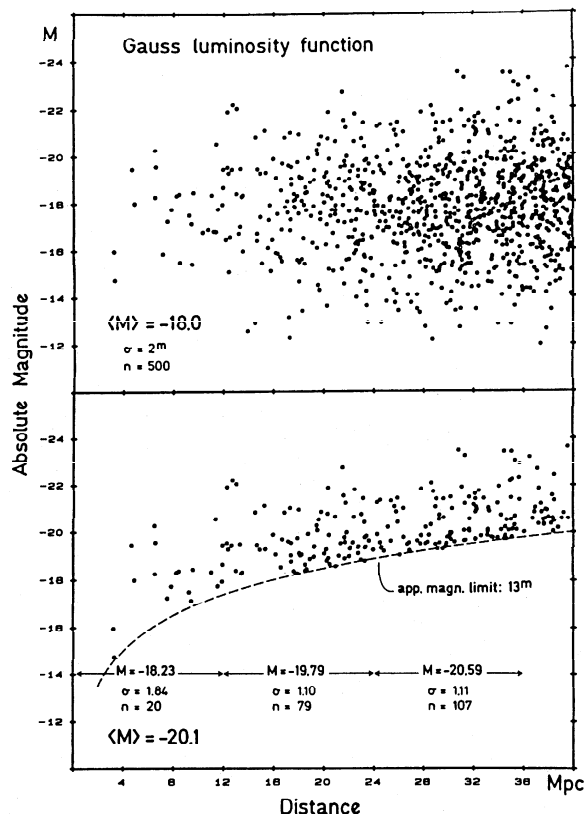,width=\floatwidth}}
\caption{{\em Upper panel:\/} Monte Carlo calculation of the distance
              and absolute magnitude of 500 galaxies with constant
              space density and $r<42\;$Mpc. Their mean absolute
              magnitude is $<\!\!M\!\!>=-18.0$ with an intrinsic
              scatter of $\sigma_M=2\fm0$.
         {\em Lower panel:\/} The same distribution cut by an
              apparent-magnitude limit $m<13.0$. Note that the mean
              luminosity of the magnitude-limited sample increases
              with distance.}
\label{fig:9}
\end{figure}

   There is an additional statistical difficulty as to the derivation
of $H_0$ from field galaxies. Frequently $H_0$ is determined from the
arithmetic mean of many values of $H_i$ found from individual field
galaxies. The underlying assumption is that the values $H_i$ have a
Gaussian distribution which is generally not correct. If the distance
errors are symmetric in the moduli $(m-M)$, they are not in linear
distance $r$ and consequently cause a skewness of $H_i$ towards high
values. An actual example is provided by the bias-corrected TF
distances of 155 field galaxies with $v<1000\kms$. A simple arithmetic
mean of their corresponding $H_i$ values, which have a non-Gaussian
distribution, would give $<\!\!H_i\!\!>=64\pm2$, while the median
value is $60.0\pm3$. Yet a more nearly correct solution is given by
averaging the values $\log H_i$, which have a perfect Gaussian
distribution. This then gives the best estimate of $H_0=58\pm2$
(Federspiel 1999).

   Unfortunately, the hope that the inverse TF relation (line width
versus magnitude) was bias-free (Sandage, Tammann, \& Yahil 1979;
Schechter 1980) has been shattered (Teerikorpi et~al.\ 1999).

   Strategies have been developed to correct for Malmquist bias,
particularly in the case of the (direct) TF relation, by e.g.,
Sandage (1988b, 1995) and Federspiel, Sandage, \& Tammann (1994) and
similarly by Teerikorpi (1994, 1997), Bottinelli et~al.\ (1986, 1995),
and Theureau (2000).

   A selection of $H_0$ values, which have been derived from
magnitude-limited, yet bias-corrected samples of field galaxies have
been compiled in Table~8. The conclusion is
\begin{equation}\label{eq:11}
   H_0=55\pm5
\end{equation}
from field galaxies, which is valid within various distance ranges up
to $\sim\!5000\kms$. It should be noted that the different methods use
spirals only -- except the D$_n-\sigma$ entry -- and their calibration
rests only on Cepheids, independent of any adopted distance of
the Virgo cluster.

\begin{table}
\footnotesize
\begin{center}
\caption{Values of $H_0$ from field galaxies corrected for Malmquist bias.}
\label{tab:8}
\begin{tabular}{llcl}
\noalign{\medskip}
  \multicolumn{1}{c}{Method} & \multicolumn{1}{c}{$H_0$} &
\multicolumn{1}{c}{Range [$\kms$]} & \multicolumn{1}{c}{Source} \\
\multicolumn{1}{c}{(1)} & \multicolumn{1}{c}{(2)} &
\multicolumn{1}{c}{(3)} & \multicolumn{1}{c}{(4)} \\
\hline
Luminosity classes              & $56\pm5$ & 4000 & Sandage 1996a \\
Morphol. twins of M\,31, M\,101 & $50\pm5$ & 5000 & Sandage 1996b \\
TF (using mag. + diam.)         & $55\pm5$ & 5000 & Theureau et~al.\ 1997 \\
Galaxy diam.                    & $50-55$  & 5000 & Goodwin et~al.\ 1997 \\
TF                              & $53\pm5$ &  500 & Federspiel 1999 \\
                                & $58\pm2$ & 1000 & $\quad\quad$ " \\
Luminosity classes              & $55\pm3$ & 4000 & Sandage 1999 \\
D$_n-\sigma$                    & $52\pm8$ & 4000 & Federspiel 1999 \\
TF (using mag. + diam.)         & $55\pm5$ & 8000 & Theureau 2000 \\
Morphol. twins of M\,31, M\,81  & $60\pm10$& 5000 & Paturel et~al.\ 1998 \\
Inverse TF                      & $53\pm6$ & 8000 & Ekholm et~al.\ 1999 \\

\end{tabular}
\end{center}
\end{table}

   Tonry et~al.\ (2000) have derived $H_0=77\pm4$ from the SBF of 300
field galaxies with $v \la 4000\kms$ and on the basis of a very
specific flow model. However, the usefulness of the SBF method has
never been tested beyond doubt; a Hubble diagram of field galaxies and
a corresponding determination of the intrinsic scatter of the method
and hence of the importance of the Malmquist bias are still
missing. The authors state correctly ``The level of this
Malmquist-like bias in our sample is difficult to quantify \dots and we
may also be subject to possible biases and selection effects which
depend on distance.'' Moreover, in discussing Table~3 it was
noted that the small relative SBF distance Virgo\,-\,Leo and the large
value Fornax\,-\,Virgo are in serious contradiction to the evidence of
Cepheids and SNe\,Ia. It is well possible that the SBF method is still
affected by hidden second-parameter effects.

\section{Conclusions}
The exceptionally tight Hubble diagram of blue SNe\,Ia out to
$\sim\!30\,000\kms$ combined with their mean absolute magnitude, which
is determined from eight galaxies whose Cepheid distances have been
determined with $HST$, offers the ideal instrument for the calibration
of the large-scale value of $H_0$. The result is -- after allowance is
made for the relatively weak dependence of SNe\,Ia luminosities on
decline rate $\Delta m_{15}$ and color and for some small systematic
effects -- $H_0=58.5\pm6.3$.

   The quoted external error is mainly determined by the calibrating
Cepheid distances. Although Cepheids are generally considered to be
the most reliable distance indicators, certain questions remain as to
their zeropoint calibration through LMC and the effect of metallicity
variations.

   Phillips et~al.\ (1999) and Suntzeff et~al.\ (1999) have adopted a
steeper relation between the SNe\,Ia luminosity and the decline rate
$\Delta m_{15}$, which is not supported by the present larger
sample. If their steep slope is taken at face value $H_0$ would be
increased by 5 percent. Gibson et~al.\ (2000) have increased the Cepheid
distances of Saha et~al.\ and Tanvir et~al.\ in Table~2 by
$0\fm11$ on average on the basis of additional Cepheids of lower
weight. This would increase $H_0$ by another 5 percent. If one 
choses to cumulate the two effects one could defend a value of $H_0=64$
(cf. Freedman 2000).

   The calibration of $H_0$ through the Hubble diagram of clusters and
through field galaxies in Section~3 and 4 has given values around
$H_0=55$, i.e. somewhat smaller, if anything, than the value from
SNe\,Ia. Since the latter offer the most direct way to $H_0$, they are
given the highest weight. A value of
\begin{equation}\label{eq:12}
   H_0 = 58 \pm 6 .
\end{equation}
is therefore adopted as the best estimate available at present.

   The two most frequent errors in the extragalactic distance scale,
which can lead to values of $H_0 > 65$, are too small a Virgo cluster
distance and Malmquist bias.

   The distance of the Virgo cluster is still sometimes equated to the
mean Cepheid distance of three Virgo spirals which are visibly on the
near side of the cluster due to their high resolution into stars, --
in fact they had been preselected on grounds of their high resolution
to facilitate work with $HST$. Thus, -- even if the larger cluster
distance of $(m-M)_{\rm Virgo}=31.60\pm0.20$ (cf. Table~4
and 5) was not required by a fourth Cepheid distance of a
bona fide cluster member (NGC\,4639) and in addition by the three
SNe\,Ia and the TF relation, -- the three Cepheid distances of highly
resolved galaxies could set only {\em a lower limit on the cluster
  distance}, i.e. $(m-M)_{\rm Virgo}>31.05$.

   The distance of the Virgo cluster has entered with high or even
full weight into the zeropoint calibration of the D$_n-\sigma$, FP,
and SBF methods and occasionally even of the turnover magnitude of the
GCLF. (The significance of the GCLF, although not yet proven beyond
doubt, is just that the calibration rests on the Cepheid distance of
M\,31 {\em and\/} on the RR~Lyr distances of Galactic globular
clusters [cf. Tammann \& Sandage 1999]). If one inserts in these cases
too low a Virgo cluster distance, one obtains incorrect values of
$H_0\approx70$, or even more, by necessity. 

   Locally calibrated distance indicators with non-negligible scatter
can exclusively be applied to distance-limited samples (or
volume-limited samples in case of clusters), but only
apparent-magnitude-limited samples are available, which frequently are
not even complete to any specific magnitude limit. The statistical
differences between a distance- and magnitude-limited sample are known
for almost eighty years (Malmquist 1920, 1922), but this insight is
violated again and again, always leading to too high values of
$H_0$. Malmquist's message is that control of the objects missing 
out to a fixed distance from a magnitude-limited catalog is equally
important as the catalog entries themselves.

   The future will see high-weight determinations of $H_0$ from
the Sunyaev-Zeldovich effect, from gravitationally lensed quasars, and
from the CMB fluctuations. At present these methods yield values of
$50\la H_0 \la 70$, which is not yet competitive with the solution of
SNe\,Ia. In the case of the CMB fluctuations one can solve for $H_0$
only in combination with other (poorly known) cosmological
parameters. Therefore an {\em independent}, high-accuracy
determination of $H_0$, to be used as a prior, is as important as
ever. 

\acknowledgments
\noindent {\bf Acknowledgement.}
The authors thank Bernd Reindl, Daniel Cerrito, and Hans Schwengeler
for their most valuable help in preparing the manuscript. G.A.T.
acknowledges financial support of the Swiss National Science Foundation.



\begin{thebibliography}{}
%
\bibitem[Alibert et~al.\ 1999]{Alibert:etal:99}
   {\sc Alibert, et~al.} 1999,
   Period-luminosity-color-radius relationships of Cepheids as a
   function of metallicity: evolutionary effects,
   {\em A\&A\/} {\bf 344}, 551.
%
\bibitem[Baum et~al.\ 1997]{Baum:etal:97}
   {\sc Baum, W.A., Hammergren, M., Thomsen, B., Groth, E.J.,
       Faber, S.M., Grillmair, C.J., \& Ajhar, E.A.} 1997,
   Distance to the Coma Cluster and a Value for Ho Inferred from
   Globular Clusters in IC 4051,
   {\em AJ\/} {\bf 113}, 1483.
%
\bibitem[Binggeli, Popescu, \& Tammann 1993]{Binggeli:etal:93}
   {\sc Binggeli, B., Popescu, C.C., \& Tammann, G.A.} 1993,
   The kinematics of the Virgo cluster revisited,
   {\em A\&AS\/} {\bf 98}, 275.
%
\bibitem[Binggeli, Sandage, \& Tammann 1985]{Binggeli:etal:85}
   {\sc Binggeli, B., Sandage, A., \& Tammann, G.A.} 1985,
   Studies of the Virgo Cluster.
   II -- A catalog of 2096 galaxies in the Virgo Cluster area,
   {\em AJ\/} {\bf 90}, 1681.
%
\bibitem[Bottinelli et~al.\ 1986]{Bottinelli:etal:86}
   {\sc Bottinelli, L., Gouguenheim, L., Paturel, G., \& Teerikorpi, P.} 1986,
   The Malmquist bias and the value of H$_0$ from the Tully-Fisher relation,
   {\em A\&A\/} {\bf 156}, 157.
%
\bibitem[Bottinelli et~al.\ 1991]{Bottinelli:etal:91} 
   {\sc Bottinelli, L., Gouguenheim, L., Paturel, G., \& Teerikorpi, P.} 1991,
   A systematic effect in the use of planetary nebulae as standard candles,
   {\em A\&A\/} {\bf 252}, 550.
%
\bibitem[Bottinelli et~al.\ 1995]{Bottinelli:etal:95}
   {\sc Bottinelli, L., Gouguenheim, L., Paturel, G., \& Teerikorpi, P.} 1995,
   Extragalactic database. VI. Inclination corrections for spiral
   galaxies and disk opaqueness in the B-band,
   {\em A\&A\/} {\bf 296}, 64.
%
\bibitem[Branch 1998]{Branch:98}
   {\sc Branch, D.} 1998,
   Type Ia Supernovae and the Hubble Constant,
   {\em ARA\&A\/} {\bf 36}, 17.
%
\bibitem[Carroll, Press, \& Turner 1992]{Carroll:etal:92}
   {\sc Carroll, S.M., Press, W.H., \& Turner, E.L.} 1992,
   The cosmological constant,
   {\em ARA\&A\/} {\bf 30}, 499.
%
\bibitem[Capaccioli et~al.\ 1989]{Capaccioli:etal:89}
   {\sc Capaccioli, M., della Valle, M., \& D'Onofrio, M.} 1989,
   Properties of the nova population in M31,
   {\em AJ\/} {\bf 97}, 1622.
%
\bibitem[Dale et~al.\ (1999)]{Dale:etal:99}
   {\sc Dale, D.A., Giovanelli, R., Haynes, M.P., Camusano, L.E., \&
   Hardy, E.} 1999,
   Seeking the Local Convergence Depth. V. Tully-Fisher Peculiar
   Velocities for 52 Abell Clusters,
   {\em AJ\/} {\bf 118}, 1489.
%
\bibitem[Dressler 1987]{Dressler:87}
   {\sc Dressler, A.} 1987,
   The D$_n-\sigma$ relation for bulges of disk galaxies -
   A new, independent measure of the Hubble constant,
   {\em ApJ\/} {\bf 317}, 1.
%
\bibitem[Ekholm et~al.\ 1999]{Ekholm:etal:99}
   {\sc Ekholm, T., Teerikorpi, P., Theureau, G., Hanski, M., Paturel,
   G., Bottinelli, L., \&  Gouguenheim, L.} 1999,
   Kinematics of the local Universe. X. H$_0$ from the inverse B-band
   Tully-Fisher relation using diameter and magnitude limited samples,
   {\em A\&A\/} {\bf 347}, 99.
%
\bibitem[Faber et~al.\ 1989]{Faber:etal:89}
   {\sc Faber, S.M., et~al.} 1989,
   Spectroscopy and photometry of elliptical galaxies.
   VI - Sample selection and data summary,
   {\em ApJS\/} {\bf 69}, 763.
%
\bibitem[Feast 1999]{Feast:99}
   {\sc Feast, M.W.} 1999,
   Cepheids as Distance Indicators,
   {\em PASP\/} {\bf 111}, 775.
%
\bibitem[Federspiel 1999]{Federspiel:99}
   {\sc Federspiel, M.} 1999,
   Kinematic Parameters of Galaxies as Distance Indicators,
   Ph.D. thesis, Univ. of Basel.
%
\bibitem[Federspiel, Sandage, \& Tammann 1994]{Federspiel:etal:94}
   {\sc Federspiel, M., Sandage, A., \& Tammann, G.A.} 1994,
   Bias properties of extragalactic distance indicators.
   III: Analysis of Tully-Fisher distances for the
   Mathewson-Ford-Buchhorn sample of 1355 galaxies,
   {\em ApJ\/} {\bf 430}, 29.
%
\bibitem[Federspiel, Tammann, \& Sandage 1998]{Federspiel:etal:98}
   {\sc Federspiel, M., Tammann, G.A., \& Sandage, A.} 1998,
   The Virgo Cluster Distance from 21 Centimeter Line Widths,
   {\em ApJ\/} {\bf 495}, 115.
%
\bibitem[Ferrarese et~al.\ 1996]{Ferrarese:etal:96}
   {\sc Ferrarese, L., et~al.} 1996,
   The Extragalactic Distance Scale Key Project. IV. The Discovery of
   Cepheids and a New Distance to M100 Using the Hubble Space
   Telescope,
   {\em ApJ\/} {\bf 464}, 568.
%
\bibitem[Ferrarese et~al.\ 2000]{Ferrarese:etal:00}
   {\sc Ferrarese, L., et~al.} 2000,
   The Hubble Space Telescope Key Project on the Extragalactic
   Distance Scale. XXVI. The Calibration of Population II Secondary
   Distance Indicators and the Value of the Hubble Constant,
   {\em ApJ\/} {\bf 529}, 745.
%
\bibitem[Freedman 2000]{Freedman:2000}
   {\sc Freedman, W.L.} 2000,
   this volume.
%
\bibitem[Garnavich et~al.\ 1996]{Garnavich:etal:96}
   {\sc Garnavich, P.M., Riess, A.G., Kirshner, R.P., Challis, P. \&
   Wagner, R.M.} 1996,
   The Spectroscopically Peculiar Supernovae 1995ac and 1995bd:
   91T and Beyond,
   {\em A\&AS\/} {\bf 189}, 4509.
%
\bibitem[Gibson et~al.\ (2000)]{Gibson:etal:00}
   {\sc Gibson, B.K., et~al.} 2000,
   The $HST$ Key Project on the Extragalactic
   Distance Scale. XXV. A Recalibration of Cepheid Distances to Type
   Ia Supernovae and the Value of the Hubble Constant,
   {\em ApJ\/} {\bf 529}, 723.
%
\bibitem[Gilmozzi \& Panagia 1999]{Gilmozzi:Panagia:99}
   {\sc Gilmozzi, R. \& Panagia, N.} 1999,
   Space Telescope Science Institute Preprint Series, No.~1319.
%
\bibitem[Giovanelli (1997)]{Giovanelli:97}
   {\sc Giovanelli, R.} 1997, private communication.
%
\bibitem[Giovanelli et~al.\ (1997)]{Giovanelli:etal:97}
   {\sc Giovanelli, R., et~al.} 1997,
   The I Band Tully-Fisher Relation for Cluster Galaxies: Data
   Presentation,
   {\em AJ\/} {\bf 113}, 22.
%
\bibitem[Goodwin et~al.\ 1997]{Goodwin:etal:97}
   {\sc Goodwin, S.P., Gribbin, J., \& Hendry, M.A.} 1997,
   A New Determination of the Hubble Parameter Using
   Galaxy Linear Diameters,
   {\em AJ\/} {\bf 114}, 2212.
%
\bibitem[Graham et~al.\ 1999]{Graham:etal:99}
   {\sc Graham, J.A., et~al.} 1999,
   The Hubble Space Telescope Key Project on the Extragalactic
   Distance Scale. XX. The Discovery of Cepheids in the Virgo Cluster
   Galaxy NGC 4548,
   {\em ApJ\/} {\bf 516}, 626.
%
\bibitem[Gratton 2000]{Gratton:00}
   {\sc Gratton, R.} 2000,
   RR Lyrae stars and Cepheids from HIPPARCOS,
   in {\em XIXth. Texas Symposium on Relativistic
   Astrophysics and Cosmology}, eds. E. Aubourg, et~al.,
   Mini-symposium 13/03.
%
\bibitem[H{\"o}flich \& Khokhlov (1996)]{Hoeflich:Khokhlov:96}
   {\sc H{\"o}flich, P.H. \& Khokhlov, A.} 1996,
   Explosion Models for Type Ia Supernovae: A Comparison with Observed
   Light Curves, Distances, H$_0$, and q$_0$,
   {\em ApJ\/} {\bf 457}, 500.
%
\bibitem[Jerjen \& Tammann (1993)]{Jerjen:Tammann:93}
   {\sc Jerjen, H., \& Tammann, G.A.} 1993,
   The Local Group Motion Towards Virgo and the Microwave Background,
   {\em A\&A\/} {\bf 276}, 1.
%
\bibitem[Jha et~al.\ (1999)]{Jha:etal:99}
   {\sc Jha, S., et~al.} 1999,
   The Type Ia Supernova 1998BU in M\,96 and the Hubble Constant,
   {\em ApJS\/} {\bf 125}, 73.
%
\bibitem[Kelson et~al.\ 2000]{Kelson:etal:00}
   {\sc Kelson, D.D., et~al.} 2000,
   The Hubble Space Telescope Key Project on the Extragalactic
   Distance Scale. XXVII. A Derivation of the Hubble Constant Using
   the Fundamental Plane and D$_{\rm n}-\sigma$ Relations in Leo I, Virgo, and
   Fornax,
   {\em ApJ\/} {\bf 529}, 768.
%
\bibitem[Kennicutt et~al.\ (1998)]{Kennicutt:etal:98}
   {\sc Kennicutt, R.C., et~al.} 1998,
   The Hubble Space Telescope Key Project on the Extragalactic
   Distance Scale. XIII. The Metallicity Dependence of the Cepheid
   Distance Scale,
   {\em ApJ\/} {\bf 498}, 181.
%
\bibitem[Kraan-Korteweg 1986]{Kraan-Korteweg:86}
   {\sc Kraan-Korteweg, R.C.} 1986,
   A catalog of 2810 nearby galaxies -- The effect of the
   Virgocentric flow model on their observed velocities,
   {\em A\&AS\/} {\bf 66}, 255.
%
\bibitem[Lanoix 1998]{Lanoix:98}
   {\sc Lanoix, P.} 1998,
   HIPPARCOS calibration of the peak brightness of four SNe\,Ia
   and the value of $H_0$,
   {\em A\&A\/} {\bf 331}, 421.
%
\bibitem[Lanoix, Paturel, \& Garnier 1999]{Lanoix:etal:99}
   {\sc Lanoix, P., Paturel, G., \& Garnier, R.} 1999,
   Bias in the Cepheid Period-Luminosity Relation,
   {\em ApJ\/} {\bf 517}, 188.
%
\bibitem[Lauer \& Postman (1992)]{Lauer:Postman:92}
   {\sc Lauer, T.R., \& Postman, M.} 1992,
   The Hubble flow from brightest cluster galaxies,
   {\em ApJ\/} {\bf 400}, L47.
%
\bibitem[Lauer et~al.\ (1998)]{Lauer:etal:98}
   {\sc Lauer, T.R., et~al.} 1998,
   The Far-Field Hubble Constant,
   {\em ApJ\/} {\bf 499}, 577.
%
\bibitem[Livio 1997]{Livio:97}
   {\sc Livio, M.} 1997,
   Novae as distance indicators,
   in {\em The Extragalactic Distance Scale},
   eds. M. Livio, M. Donahue, \& N. Panagia,
   (Cambridge: Cambridge Univ. Press), p.~186.
%
\bibitem[Madore \& Freedman 1998]{Madore:Freedman:98}
   {\sc Madore, B., \& Freedman, W.L.} 1998,
   Hipparcos Parallaxes and the Cepheid Distance Scale,
   {\em ApJ\/} {\bf 492}, 110.
%
\bibitem[Macri et~al.\ 1999]{Macri:etal:99}
   {\sc Macri, L.M., et~al.} 1999,
   The Extragalactic Distance Scale Key Project. XVIII. The Discovery of
   Cepheids and a New Distance to NGC 4535 Using the Hubble Space
   Telescope,
   {\em ApJ\/} {\bf 521}, 155.
%
\bibitem[Malmquist (1920)]{Malmquist:20}
   {\sc Malmquist, K.G.} 1920,
   A study of the stars of spectral type A,
   {\em Lund Medd. Ser.~II\/} {\bf 22}, 1.
%
\bibitem[Malmquist (1922)]{Malmquist:22}
   {\sc Malmquist, K.G.} 1922,
   On some relations in stellar statistics,
   {\em Lund Medd. Ser.~I\/} {\bf 100}, 1.
%
\bibitem[Mazumdar \& Narasimha 2000]{Mazumdar:Narasimha:00}
   {\sc Mazumdar, A. \& Narasimha, D.} 2000,
   Calibration of Cepheid characteristics relevant to the distance
   scale, 
   preprint.
%
\bibitem[M{\'e}ndez et~al. 1993]{Mendez:etal:93}
   {\sc M{\'e}ndez, R.H., Kudritzki, R.P., Ciardullo, R., \& Jacoby,
   G.H.} 1993,
   The bright end of the planetary nebula luminosity function,
   {\em A\&A\/} {\bf 275}, 534.
%
\bibitem[Mould et~al.\ 2000]{Mould:etal:00}
   {\sc Mould, J.R., et~al.} 2000,
   The Hubble Space Telescope Key Project on the Extragalactic
   Distance Scale. XXI. The Cepheid Distance to NGC 1425,
   {\em ApJ\/} {\bf 528}, 655.
%
\bibitem[Parodi et~al.\ 2000]{Parodi:etal:00}
   {\sc Parodi, B.R., Saha, A., Sandage, A., \& Tammann, G.A.} 2000,
   Supernova Type Ia Luminosities and Their Dependence on Second
   Parameters, {\em ApJ}, in press.
%
\bibitem[Paturel et~al.\ 1998]{Paturel:etal:98}
   {\sc Paturel, et~al.} 1998,
   Hubble constant from sosie galaxies and HIPPARCOS geometrical
   calibration, 
   {\em A\&A\/} {\bf 339}, 671.
%
\bibitem[Perlmutter 1998]{Perlmutter:98}
   {\sc Perlmutter, S.} 1998,
   Measurements of $\Omega$ and $\Lambda$ from Supernovae,
   in {\em Supernovae and Cosmology}, eds. L.~Labhardt, B.~Binggeli,
   \& R.~Buser, (Basel: Astronomisches Institut der Universit{\"a}t
   Basel), p.~75.
%
\bibitem[Perlmutter et~al.\ 1999]{Perlmutter:etal:99}
   {\sc Perlmutter, S., et~al.} 1999,
   Measurements of Omega and Lambda from 42 High-Redshift Supernovae,
   {\em ApJ\/} {\bf 517}, 565.
%
\bibitem[Philips (1993)]{Phillips:93}
   {\sc Phillips, M.M.} 1993,
   The absolute magnitudes of Type Ia supernovae,
   {\em ApJ\/} {\bf 413}, 105.
%
\bibitem[Phillips et~al.\ 1992]{Phillips:etal:92}
   {\sc Phillips, M. M., Wells, L.A., Suntzeff, N.B., Hamuy, M.,
   Leibundgut, B., Kirshner, R.P., \& Foltz, C.B.} 1992,
   SN 1991T - Further evidence of the heterogeneous nature of
   type Ia supernovae, {\em AJ\/} {\bf 103}, 1632.
%
\bibitem[Phillips et~al.\ (1999)]{Phillips:etal:99}
   {\sc Phillips, M.M., Lira, P., Suntzeff, N.B., Schommer R.A.,
   Hamuy, M., \& Maza, J.} 1999,
   The Reddening-Free Decline Rate Versus Luminosity Relationship
   for Type Ia Supernovae,
   {\em AJ\/} {\bf 118}, 1766.
%
\bibitem[Prosser et~al.\ 1999]{Prosser:etal:99}
   {\sc Prosser, C.F., et~al.} 1999,
   The Hubble Space Telescope Key Project on the Extragalactic
   Distance Scale. XXII. The Discovery of Cepheids in NGC 1326A,
   {\em ApJ\/} {\bf 525}, 80.
%
\bibitem[Pritchet \& van den Bergh]{Pritchet:vandenBergh:87}
   {\sc Pritchet, C.J., \& van den Bergh, S.} 1987,
   Observations of novae in the Virgo cluster,
   {\em ApJ\/} {\bf 318}, 507.
%
\bibitem[Richtler \& Drenkhahn 1999]{Richtler:Drenkhahn:99}
   {\sc Richtler, T., \& Drenkhahn, G.} 1999,
   The Hubble Constant from Type Ia Supernovae in Early-Type Galaxies,
   in {\em Cosmology and Astrophysics: A collection of critical
           thoughts},
   eds. W. Kundt \&  C. van de Bruck,
   Lecture Notes in Physics, (Berlin: Springer), astro-ph/99\,09\,117.
%
\bibitem[Riess et~al.\ 1998]{Riess:etal:98}
   {\sc Riess, A.G., et~al.} 1998,
   Observational Evidence from Supernovae for an Accelerating Universe
   and a Cosmological Constant,
   {\em AJ\/} {\bf 116}, 1009.
%
\bibitem[Saha et~al.\ (1994)]{Saha:etal:94}
   {\sc Saha, A., Labhardt, L., Schwengeler, H., Macchetto, F.D.,
   Panagia, N., Sandage, A., \& Tammann, G.A.} 1994,
   Discovery of Cepheids in IC\,4182: Absolute peak brightness of
   SN\,Ia 1937C and the value of H$_0$,
   {\em ApJ\/} {\bf 425}, 14.
%
\bibitem[Saha et~al.\ (1995)]{Saha:etal:95}
   {\sc Saha, A., Sandage, A., Labhardt, L., Schwengeler, H., Tammann,
   G.A., Panagia, N., \& Macchetto, F.D.} 1995,
   Discovery of Cepheids in NGC\,5253: Absolute peak brightness of
   SN\,Ia 1895B and SN\,Ia 1972E and the value of H$_0$,
   {\em ApJ\/} {\bf 438}, 8.
%
\bibitem[Saha et~al.\ (1996a)]{Saha:etal:96a}
   {\sc Saha, A., Sandage, A., Labhardt, L., Tammann, G.A., Macchetto,
   F.D., \& Panagia, N.} 1996a,
   Cepheid Calibration of the Peak Brightness of SNe\,Ia.
   V. SN\,1981B in NGC\,4536,
   {\em ApJ\/} {\bf 466}, 55.
%
\bibitem[Saha et~al.\ (1996b)]{Saha:etal:96b}
   {\sc Saha, A., Sandage, A., Labhardt, L., Tammann, G.A., Macchetto,
   F.D., \& Panagia, N.} 1996b,
   Cepheid Calibration of the Peak Brightness of SNe\,Ia.
   V. SN\,1981B in NGC\,4536,
   {\em ApJS\/} {\bf 107}, 693.
%
\bibitem[Saha et~al.\ (1997)]{Saha:etal:97}
   {\sc Saha, A., Sandage, A., Labhardt, L., Tammann, G.A., Macchetto,
   F.D., \& Panagia, N.} 1997,
   Cepheid Calibration of the Peak Brightness of Type Ia Supernovae.
   VIII. SN 1990N in NGC 4639,
   {\em ApJ\/} {\bf 486}, 1.
%
\bibitem[Saha et~al.\ 1999]{Saha:etal:99}
   {\sc Saha, A., Sandage, A., Tammann, G.A., Labhardt, L., Macchetto,
   F.D., \& Panagia, N.} 1999,
   Cepheid Calibration of the Peak Brightness of Type Ia Supernovae.
   IX. SN 1989B in NGC 3627,
   {\em ApJ\/} {\bf 522}, 802.
%
\bibitem[Saha et~al.\ 2000a]{Saha:etal:00a}
  {\sc Saha, A., Sandage, A., Thim, F., Labhardt, L., Tammann, G.A.,
  Macchetto, F.D., \& Panagia, N.} 2000a,
  Cepheid Calibration of the Peak Brightness of SNe Ia. X. SN~1991T in
  NGC~4527, {\em ApJ}, in press.
%
\bibitem[Saha et~al.\ 2000b]{Saha:etal:00b}
  {\sc Saha, A., Labhardt, L., \& Prosser, C.} 2000b,
  On Deriving Distances from Cepheids Using the Hubble Space Telescope,
  {\em PASP\/} {\bf 112}, 163.
%
\bibitem[Sakai, Zwitsky, \& Kennicutt 2000]{Sakai:etal:00a}
   {\sc Sakai, S., Zwitsky, D., \& Kennicutt, R.C.} 2000,
   The Tip of the Red Giant Branch Distance to the Large Magellanic
   Cloud,
   {\em AJ\/} {\bf 119}, 1197.
%
\bibitem[Sakai et~al.\ 2000]{Sakai:etal:00}
  {\sc Sakai, S., et~al.} 2000,
  The Hubble Space Telescope Key Project on the Extragalactic Distance
  Scale. XXIV. The Calibration of Tully-Fisher Relations and the Value
  of the Hubble Constant,
  {\em ApJ\/} {\bf 529}, 698. 
%
\bibitem[Sandage (1961)]{Sandage:61}
   {\sc Sandage, A.} 1961,
   The Ability of the 200-inch Telescope to Discriminate Between
   Selected World Models,
   {\em ApJ\/} {\bf 133}, 355.
%
\bibitem[Sandage (1962)]{Sandage:62}
   {\sc Sandage, A.} 1962,
   The Change of Redshift and Apparent Luminosity of Galaxies due to
   the Deceleration of Selected Expanding Universes,
   {\em ApJ\/} {\bf 136}, 319.
%
\bibitem[Sandage (1988)]{Sandage:88a}
   {\sc Sandage, A.} 1988,
   Cepheids as distance indicators when used near their
   detection limit,
   {\em PASP\/} {\bf 100}, 935.
%
\bibitem[Sandage 1988]{Sandage:88b}
   {\sc Sandage, A.} 1988,
   A case for $H_0 = 42$ and ${\Omega}_{0} = 1$ using luminous spiral
   galaxies and the cosmological time scale test,
   {\em ApJ\/} {\bf 331}, 583.
%
\bibitem[Sandage 1995]{Sandage:95}
   {\sc Sandage, A.} 1995,
   Practical Cosmology: Inverting the Past,
   in {\em The Deep Universe}, eds. B. Binggeli \& R. Buser,
   (Springer: Berlin), p.~210.
%
\bibitem[Sandage 1996a]{Sandage:96a}
   {\sc Sandage, A.} 1996a,
   Bias Properties of Extragalactic Distance Indicators.
   V. $H_0$ From Luminosity Functions of Different Spiral
   Types and Luminosity Classes Corrected for Bias,
   {\em AJ\/} {\bf 111}, 1.
%
\bibitem[Sandage 1996b]{Sandage:96b}
   {\sc Sandage, A.} 1996b,
   Bias Properties of Extragalactic Distance Indicators.
   VI. Luminosity Functions of M\,31 and M\,101
   Look-alikes Listed in the RSA2: $H_0$ Therefrom,
   {\em AJ\/} {\bf 111}, 18.
%
\bibitem[Sandage 1999]{Sandage:99}
   {\sc Sandage, A.} 1999,
   Bias Properties of Extragalactic Distance Indicators. VIII. H0 from
   Distance-limited Luminosity Class and Morphological Type-Specific
   Luminosity Functions for Sb, Sbc, and Sc Galaxies Calibrated Using
   Cepheids,
   {\em ApJ\/} {\bf 527}, 479.
%
\bibitem[Sandage \& Bedke (1988)]{Sandage:Bedke:88}
   {\sc Sandage, A., \& Bedke, J.} 1988, 
   {\em Atlas of Galaxies useful to measure the Cosmological Distance
   Scale}, (NASA: Washington).
%
\bibitem[Sandage, Bell, \& Tripicco 1999]{Sandage:etal:99}
   {\sc Sandage, A., Bell, R.A., \& Tripicco, M.J.} 1999,
   On the Sensitivity of the Cepheid Period-Luminosity Relation to
   Variations of Metallicity,
   {\em ApJ\/} {\bf 522}, 250 .
%
\bibitem[Sandage \& Hardy 1973]{Sandage:Hardy:73}
   {\sc Sandage, A., \& Hardy, E.} 1973,
   The Redshift-Distance Relation. VIL Absolute Magnitudes of the
   First Three Ranked Cluster Galaxies as Functions of Cluster
   Richness and Bautz-Morgan Cluster Type: the Effect of q$_{0}$,
   {\em ApJ\/} {\bf 183}, 743.
%
\bibitem[Sandage, Tammann, \& Federspiel 1995]{Sandage:etal:95}
   {\sc Sandage, A., Tammann, G.A., \& Federspiel, M.} 1995,
   Bias Properties of Extragalactic Distance Indicators.
   IV. Demonstration of the Population Incompleteness Bias Inherent in
   the Tully-Fisher Method Applied to Clusters,
   {\em ApJ\/} {\bf 452}, 1.
%
\bibitem[Sandage, Tammann, \& Yahil 1979]{Sandage:etal:79}
   {\sc Sandage, A., Tammann, G.A., \& Yahil, A.} 1979,
   The velocity field of bright nearby galaxies. I - The variation of
   mean absolute magnitude with redshift for galaxies in a
   magnitude-limited sample,
   {\em ApJ\/} {\bf 232}, 352.
%
\bibitem[Schechter 1980]{Schechter:80}
   {\sc Schechter, P.L.} 1980, Mass-to-light ratios for elliptical
   galaxies, 
   {\em AJ\/} {\bf 85}, 801.
%
\bibitem[Schlegel et~al. (1998)]{Schlegel:etal:98}
   {\sc Schlegel, D., Finkbeiner, D., \& Davis, M.} 1998,
   Maps of Dust Infrared Emission for Use in Estimation of Reddening
   and Cosmic Microwave Background Radiation Foregrounds,
   {\em ApJ\/} {\bf 500}, 525.
%
\bibitem[Schmidt et~al. 1998]{Schmidt:etal:98}
   {\sc Schmidt, B., et~al.} 1998,
   The High-Z Supernova Search: Measuring Cosmic Deceleration
   and Global Curvature of the Universe Using Type Ia Supernovae,
   {\em ApJ\/} {\bf 507}, 46.
%
\bibitem[Schr{\"o}der 1995]{Schroeder:95}
   {\sc Schr{\"o}der, A.} 1995,
   Ph.D. thesis, Univ. of Basel.
%
\bibitem[Silbermann et~al. 1999]{Silbermann:etal:99}
   {\sc Silbermann, N.A., et~al.} 1999,
   The Hubble Space Telescope Key Project on the Extragalactic
   Distance Scale. XIV. The Cepheids in NGC 1365,
   {\em ApJ\/} {\bf 515}, 1.
%
\bibitem[Soffner et~al. 1996]{Soffner:etal:96}
   {\sc Soffner, T., M{\'e}ndez, R., Jacoby, G., Ciardullo, R., Roth, M.,
   \& Kudritzki, R.} 1996,
   Planetary nebulae and HII regions in NGC 300,
   {\em A\&A\/} {\bf 306}, 9.
%
\bibitem[Stanek \& Udalski (1999)]{Stanek:Udalski:99}
   {\sc  Stanek, K.Z., \& Udalski, A.} 1999,
   The Optical Gravitational Lensing Experiment. Investigating the
   Influence of Blending on the Cepheid Distance Scale with Cepheids
   in the Large Magellanic Cloud, preprint, astro-ph/99\,09\,346 .
%
\bibitem[Suntzeff et~al.\ (1999)]{Suntzeff:etal:99}
   {\sc Suntzeff, N.B., et~al.} 1999,
   Optical Light Curve of the Type Ia Supernova 1998BU in M\,96 and the
   Supernova Calibration of the Hubble Constant,
   {\em AJ\/} {\bf 117}, 1175.
%
\bibitem[Tammann 1982]{Tammann:82}
   {\sc Tammann, G.A.} 1982,
   Supernova statistics and related problems,
   in {\em Supernovae: A Survey of Current Research},
   eds. M.J. Rees \& R.J. Stoneham, (Dordrecht: Reidel), p.~371.
%
\bibitem[Tammann 1993]{Tammann:93}
   {\sc Tammann, G.A.} 1993,
   Why are Planetary Nebulae Poor Distance Indicators?
   in {\em Planetary Nebulae}, eds. R.~Weinberger \&  A.~Acker,
   IAU Symp.~155, (Dordrecht: Kluwer), p.~515.
%
\bibitem[Tammann \& Federspiel 1997]{Tammann:Federspiel:97}
   {\sc Tammann, G.A., \& Federspiel, M.} 1997,
   Focusing in on $H_0$,
   in {\em The Extragalactic Distance Scale},
   eds. M. Livio, M. Donahue, \& N. Panagia,
   (Cambridge: Cambridge Univ. Press), p.~137.
%
\bibitem[Tammann \& Sandage 1999]{Tammann:Sandage:99}
   {\sc Tammann, G.A., \& Sandage, A.} 1999,
   The Luminosity Function of Globular Clusters as an Extragalactic
   Distance Indicator,
   in {\em Harmonizing Cosmic
   Distance Scales in a Post-Hipparcos Era}, eds. D. Egret \& A. Heck,
   p.~204.
%
\bibitem[Tammann, Sandage, \& Reindl 2000]{Tammann:etal:00}
   {\sc Tammann, G.A., Sandage, A., \& Reindl, B.} 2000,
   The Distance of the  Virgo Cluster,
   in {\em XIXth. Texas Symposium on Relativistic
   Astrophysics and Cosmology}, eds. E. Aubourg, et~al.,
   Mini-Symposium 13/11.
%
\bibitem[Tanvir et~al.\ (1995)]{Tanvir:etal:95}
  {\sc Tanvir, N.R., Shanks, T., Ferguson, H.C., \& Robinson, D.R.T.}
  1995, Determination of the Hubble constant from observations of
  Cepheid variables in the galaxy M\,96,
  {\em Nature\/} {\bf 377}, 27.
%
\bibitem[Teerikorpi 1984]{Teerikorpi:84}
   {\sc Teerikorpi, P.} 1984,
   Malmquist bias in a relation of the form M = aP + b,
   {\em A\&A\/} {\bf 141}, 407.
%
\bibitem[Teerikorpi 1997]{Teerikorpi:97}
   {\sc Teerikorpi, P.} 1997,
   Observational Selection Bias Affecting the Determination of the
   Extragalactic Distance Scale,
   {\em ARA\&A\/} {\bf 35}, 101.
%
\bibitem[Teerikorpi et~al.\ 1999]{Teerikorpi:etal:99}
   {\sc Teerikorpi, P., Ekholm, T., Hanski, M.O., \& Theureau, G.} 1999,
   Theoretical aspects of the inverse Tully-Fisher relation as a
   distance indicator: incompleteness in $\log V_{\rm Max}$, the relevant
   slope, and the calibrator sample bias,
   {\em A\&A\/} {\bf 343}, 713.
%
\bibitem[Theureau (2000)]{Theureau:00}
   {\sc Theureau, G.} 2000,
   Tully-Fisher distances of field galaxies and the value of $H_0$,
   in {\em XIXth. Texas Symposium on Relativistic
   Astrophysics and Cosmology}, eds. E. Aubourg, et~al.,
   Mini-Symposium 13/12.
%
\bibitem[Theureau et~al.\ 1997]{Theureau:etal:97}
   {\sc Theureau, G., Hanski, M., Ekholm, T., Bottinelli, L.,
   Gouguenheim, L., Paturel, G., \& Teerikorpi, P.} 1997,
   Kinematics of the Local Universe. V. The value of $H_0$ from the
   Tully-Fisher B and  $\log D_{25}$ relations for field galaxies,
   {\em A\&A\/} {\bf 322}, 730.
%
\bibitem[Tonry et~al.\ 2000]{Tonry:etal:00}
   {\sc Tonry, J.L., Blakeslee, J.P., Ajhar, E.A., \& Dressler, A.} 2000,
   The Surface Brightness Fluctuation Survey of Galaxy
   Distances. II. Local and Large-Scale Flows,
   {\em ApJ\/} {\bf 530}, 625.
%
\bibitem[Tripp 1998]{Tripp:98}
   {\sc Tripp, R.} 1998,
   A two-parameter luminosity correction for Type Ia supernovae,
   {\em A\&A\/} {\bf 331}, 815.
%
\bibitem[Tripp and Branch (1999)]{Tripp:Branch:99}
   {\sc Tripp, R., \& Branch, D.} 1999,
   Determination of the Hubble Constant Using a Two-Parameter
   Luminosity Correction for Type Ia Supernovae,
   {\em ApJ\/} {\bf 525}, 209.
%
\bibitem[Turner et~al.\ 1998]{Turner:etal:98}
   {\sc Turner, A., et~al.} 1998,
   The Hubble Space Telescope Key Project on the Extragalactic
   Distance Scale. XI. The Cepheids in NGC\,4414,
   {\em ApJ\/} {\bf 505}, 207.
%
\bibitem[Walker 1999]{Walker:99}
   {\sc Walker, A.R.} 1999,
   The Distances of the Magellanic Clouds,
   in {\em Post-Hipparcos cosmic candles}, eds. A. Heck and
   F. Caputo, (Dordrecht: Kluwer Academic Publishers), p.~125.
%
\bibitem[Weedman (1976)]{Weedman:76}
   {\sc Weedman, D.} 1976,
   The Hubble diagram for nuclear magnitudes of cluster galaxies,
   {\em ApJ\/} {\bf 203}, 6.
%
\end{thebibliography}
\end{document}